\documentclass[11pt]{article}
\usepackage[centertags]{amsmath}
\hsize \textwidth
\advance \hsize by -\marginparwidth
\evensidemargin
\oddsidemargin
\usepackage{amssymb}
\usepackage{graphicx}

\advance\hoffset by 5mm

\usepackage{amssymb}

\newtheorem{df}{Definition}[section]
\newtheorem{lemma}[df]{Lemma}
\newtheorem{prop}[df]{Proposition}
\newtheorem{thm}[df]{Theorem}
\newtheorem{e}[df]{Example}
\newtheorem{cor}[df]{Corollary}

\newcommand{\pdr}[2]{\dfrac{\partial{#1}}{\partial{#2}}}

\makeatletter \@addtoreset{equation}{section}

\newcommand{\bes}{\begin{displaymath}}
\newcommand{\ees}{\end{displaymath}}
\newcommand{\be}{\begin{equation}}
\newcommand{\ee}{\end{equation}}
\newcommand{\ba}{\begin{eqnarray}}
\newcommand{\ea}{\end{eqnarray}}
\newcommand{\bas}{\begin{eqnarray*}}
\newcommand{\eas}{\end{eqnarray*}}
\newcommand{\@Bbb}[1]{\ensuremath{\mathbb #1}}

\newcommand{\B}{{\@Bbb B}}
\newcommand{\C}{{\@Bbb C}}
\newcommand{\E}{{\@Bbb E}}
\newcommand{\F}{{\@Bbb F}}
\renewcommand{\P}{{\@Bbb P}}

\newcommand{\Q}{{\@Bbb Q}}
\newcommand{\bQ}{{\@Bbb Q}}
\newcommand{\N}{{\@Bbb N}}
\newcommand{\R}{{\@Bbb R}}
\newcommand{\bbR}{{\@Bbb R}}
\newcommand{\W}{{\@Bbb W}}
\newcommand{\Z}{{\@Bbb Z}}
\newcommand{\bbZ}{{\@Bbb Z}}
\newcommand{\bX}{\mathbf X}
\newcommand{\vx}{\mathbf x}

\newcommand{\bbe}{\mbox{\bf e}}
\newcommand{\Rm}{{\mathbb R}}

\newcommand{\la}{\lambda}
\newcommand{\al}{\alpha}
\newcommand{\bt}{\beta}

\newcommand{\bone}{{\bf 1}}
\newcommand{\Om}{\Omega}
\newcommand{\om}{\omega}
\newcommand{\vv}{{\bf v}}
\newcommand{\vV}{{\bf V}}
\newcommand{\bk}{{\bf k}}

\newcommand{\bp}{{\bf p}}
\newcommand{\bze}{{\bf 0}}

\newcommand{\bB}{{\bf B}}
\newcommand{\bD}{{\bf D}}
\newcommand{\bF}{{\bf F}}
\newcommand{\vF}{{\bf F}}
\newcommand{\bG}{{\bf G}}

\newcommand{\bR}{{\bf R}}

\newcommand{\bz}{{\bf z}}
\newcommand{\@s}[1]{\ensuremath{\mathcal #1}}
\newcommand{\cA}{\@s A}
\newcommand{\cB}{\@s B}
\newcommand{\cC}{\@s C}
\newcommand{\cD}{\@s D}
\newcommand{\cE}{\@s E}
\newcommand{\cF}{\@s F}
\newcommand{\cG}{\@s G}
\newcommand{\cH}{\@s H}
\newcommand{\cI}{\@s I}
\newcommand{\cJ}{\@s J}
\newcommand{\cK}{\@s K}
\newcommand{\cL}{\@s L}
\newcommand{\cN}{\@s N}
\newcommand{\cM}{\@s M}
\newcommand{\cO}{\@s O}
\newcommand{\cP}{\@s P}
\newcommand{\cR}{\@s R}
\newcommand{\cS}{\@s S}
\newcommand{\cT}{\@s T}
\newcommand{\cV}{\@s V}
\newcommand{\cW}{\@s W}
\newcommand{\cX}{\@s X}
\newcommand{\cY}{\@s Y}
\newcommand{\cZ}{\@s Z}

\newcommand{\@bm}[1]{\ensuremath{\mathbf #1}}
\newcommand{\bma}{\@bm a}
\newcommand{\bmb}{\@bm b}
\newcommand{\bmc}{\@bm c}
\newcommand{\bmd}{\@bm d}

\newcommand{\bme}{\@bm e}
\newcommand{\bmf}{\@bm f}
\newcommand{\bmg}{\@bm g}
\newcommand{\bmh}{\@bm h}
\newcommand{\bmi}{\@bm i}
\newcommand{\bmj}{\@bm j}
\newcommand{\bmk}{\@bm k}
\newcommand{\bml}{\@bm l}
\newcommand{\bmm}{\@bm m}
\newcommand{\bmn}{\@bm n}
\newcommand{\bmo}{\@bm o}
\newcommand{\bmp}{\@bm p}
\newcommand{\bmq}{\@bm q}
\newcommand{\bmr}{\@bm r}
\newcommand{\bms}{\@bm s}
\newcommand{\bmt}{\@bm t}
\newcommand{\bmu}{\@bm u}
\newcommand{\bmw}{\@bm w}
\newcommand{\bmv}{\@bm v}
\newcommand{\bmx}{\@bm x}
\newcommand{\bx}{\@bm x}
\newcommand{\bmy}{\@bm y}
\newcommand{\bmz}{\@bm z}

\newcommand{\by}{\@bm y}
\newcommand{\bmzero}{\@bm 0}

\newcommand{\@g}[1]{\ensuremath{\mathfrak #1}}
\newcommand{\gA}{\@g A}
\newcommand{\gD}{\@g D}
\newcommand{\gJ}{\@g J}
\newcommand{\gF}{\@g F}
\newcommand{\gM}{\@g M}
\newcommand{\gR}{\@g R}
\newcommand{\ie}{{\mbox{e}}}

\newcommand{\commentout}[1]{{}}

\makeatother
\begin{document}

\title{
Passive tracer in a slowly decorrelating random flow with a large
mean}
\author{Tomasz Komorowski
\thanks{Institute of Mathematics, UMCS, pl. Marii Curie-Sk\l odowskiej 1,
Lublin 20-031, Poland; \newline\rm {\texttt{
komorow@hektor.umcs.lublin.pl}}}
\and Lenya Ryzhik\thanks{Department of Mathematics, University of Chicago,
Chicago, IL 60637, USA; \rm {\texttt{
ryzhik@math.uchicago.edu}}}}

\maketitle

\begin{abstract}
We consider the movement of a particle advected by a random flow of
the form $\vv+\delta \bF(\vx)$, with $\vv\in\R^d$ a constant drift,
$\bF(\vx)$ -- the fluctuation -- given by a zero mean, stationary
random field and $\delta\ll 1$ so that the drift
dominates over the fluctuation. The two-point correlation matrix
$\bR(\vx)$ of the random field decays as $|\vx|^{2\alpha-2}$, as $|\vx|\to+\infty$
 with
$\alpha<1$. The Kubo formula for the
effective diffusion coefficient obtained in \cite{kp79} for rapidly
decorrelating fields diverges when $1/2\le\alpha<1$. We show
formally that on the time scale $\delta^{-1/\alpha}$ the deviation
of the trajectory from its mean $\by(t)=\vx(t)-\vv t$ converges to a
fractional Brownian motion $B_\alpha(t)$ in this range of the
exponent $\alpha$. We also prove rigorously upper and lower bounds
which show that $\E[|\by(t)|^2]$ converges to zero for times
$t\ll\delta^{-1/\alpha}$ and to infinity on time scales $t\gg
\delta^{-1/\alpha}$ as $\delta\to 0$ when $\alpha\in(1/2,1)$. On the
other hand, when $\alpha<1/2$ non-trivial behavior is observed on
the  time-scale $O(\delta^{-2})$.
\end{abstract}

\section{Introduction}
\label{intro}
The position of a tracer in a random flow is described by the ordinary differential equation
\begin{equation}\label{basic}
     \frac{d\bX(t;\bx)}{dt}=\vV(t,\bX(t;\bx)),\quad \bX(0;\bx)=\bx,
\end{equation}
where $\vV(t,\bx)$ is a random field.  This random model is
frequently used to describe the motion of a particle in a turbulent
flow of fluid, where $\vV(t,\bmx)$ is the Eulerian velocity field.
The long-time behavior of solutions of (\ref{basic}) has been
extensively studied, especially  when $\vV(t,\bx)$ is sufficiently
rapidly mixing in time and space -- see \cite{kramer-majda} for an
extensive overview.  Roughly speaking, one expects that the trajectory
$\bX(t;\bx)$ behaves diffusively if the velocity field decorrelates
sufficiently fast in time and space while an anomalous behavior is
observed if $\vV(t,\vx)$ has long range correlations. This problem
remains largely open when $\vV(\bx)$ is both time-independent and
of zero mean, as then there is no mechanism to ``move the particle
around'' and employ the decorrelating properties of the random
field. Therefore, to simplify matters, we assume that the flow
satisfies \emph{the Taylor hypothesis}, that is, it has a non-zero
mean drift which dominates over the amplitude of its fluctuations.
This means that we can write $\vV(\bx)=\vv+\delta\bF(\bx)$, where
$\vv\not=\bze$ is a constant vector with, say, $|\vv|=1$, and the
parameter $\delta\ll1$. The fluctuation $\bF(\bx)$ is spatially
homogeneous, of zero mean and is divergence-free:
$\nabla\cdot\bF(\bx)=0$. It has been shown in \cite{kp79} that if the
field $\bF(\bx)$ is sufficiently strongly mixing in space then the
scaled trajectories
\begin{equation}\label{y-proc}
\bmy(t/\delta^2;\bx):=\bmx(t/\delta^2;\bx)-\vv t/\delta^2, \quad t\ge0
\end{equation}
converge, as $\delta\downarrow 0$, as continuous stochastic processes, to a zero mean Brownian motion
starting at $\bx$ with the covariance matrix $\bD=[D_{i,j}]$  given by
the Kubo-Taylor formula
\begin{equation}\label{kubo-form}
D_{ij}=\frac12\int\limits_0^{+\infty}\left[R_{ij}(\vv t)+R_{ji}(\vv t)\right]dt,\quad i,j=,\ldots,d.
\end{equation}
Here $\bR(\bx)=[R_{ij}(\bx)]$ is the covariance matrix
of the field $\bF(\bx)$, that is, $R_{ij}(\bx)=\E[F_i(\bx)F_j(\bze)]$.

In the present paper we are interested in the situation when the
correlation tensor decays slowly in space so that the diffusion matrix
given by the Kubo-Taylor formula is infinite and the standard
diffusion limit may fail. More precisely, we consider Gaussian fields
which are isotropic and satisfy locally the self-similarity
hypothesis. This  means that the two-point correlation tensor
satisfies
\begin{equation}\label{covariance}
R_{ij}(\bmx)=
\int_{\R^d}\ie^{i\bk\cdot\bx}\hat{R}_{ij}(\bk)d\bk,
\end{equation}
with the power spectrum
\begin{equation}\label{pow-energy}
\hat{R}_{ij}(\bk)=\frac{a(|\bk|)}{|\bk|^{2\al+d-2}}\Gamma_{ij}(\hat\bk),\quad i,j=,\ldots,d.
\end{equation}
Here $a(|\bk|)\ge 0$ is a non-negative bounded, measurable function,
supported in a finite ball  $\{|\bk|\le K\}$ for some $K>0$,
continuous at $\bk=0$ and  with $a(0)=1$. The factor
$\Gamma_{ij}(\hat\bk):=\delta_{ij}-\hat k_i\hat k_j$, where
$\hat\bk=(\hat k_1,\ldots,\hat k_d):=\bk/|\bk|$, ensures
incompressibility of the flow. To guarantee the integrability of the
spectrum we assume that $\al<1$.  The rate of decay of the
correlations of the field is then given by
\begin{equation}\label{corr}
R_{ij}(\bx)\sim |\bx|^{2\al-2},\quad \mbox{for }|\bx|\gg1.
\end{equation}
A simple calculation shows that in this situation the diffusion matrix given by
the Kubo formula (\ref{kubo-form}) is infinite, provided that
$1/2<\alpha<1$. The goal of this paper is to find the proper time
scaling for the tracer trajectory when the Kubo-Taylor formula
diverges.  In other words we are looking
 for $H$ such that for any $\rho>0$
\begin{equation}\label{H-scaling}
\lim_{\delta\to0+}\mathbb E|\bmy(t/\delta^{2H(1+\rho)};\bx)|^2=+\infty
~~~~~~\hbox{ and }~~~~~~~
 \lim_{\delta\to0+}
\mathbb E|\bmy(t/\delta^{2H(1-\rho)};\bx)|^2=0.
\end{equation}
The result of Kesten and Papanicolaou, see Theorem 4 of \cite{kp79},
guarantees that for $\bR(\vx)$, which is decaying sufficiently fast, or,
equivalently, for $\al<-N$ with a large enough $N>0$ one has $H=1$.
Note, however, that the Kubo--Taylor formula~\eqref{kubo-form} itself
makes sense for any $\al<1/2$, that is, in a much larger range of
$\alpha$. We shall prove rigorously in Section \ref{sec:main-results}
below that then indeed $H=1$ for $\alpha<1/2$, with a slightly
modified definition of the limit appearing in
condition~\eqref{H-scaling}.

According to \eqref{corr}, increasing  the value of $\al$ leads to 
strengthening of the correlations of the corresponding field (and
degraded mixing properties). We shall prove, see Theorems \ref{main1} and \ref{main2}
below, that the level $\al=1/2$ is critical in the sense that the
value of the exponent $H$ changes at $\alpha=1/2$ from $H=1/2$ for $\alpha<1/2$ to
$H=1/(2\al)<1$ for $1/2<\alpha<1$.

We also present a formal argument which provides more refined
information on the process $\by(t)$ for $\alpha\in(1/2,1)$. In
addition to identifying the time scale $t\sim O(\delta^{-2H})$, $H=1/(2\alpha)$
on which $\by(t)$ behaves non-trivially,
we are able to compute formally the limit of the process
\[
\bmy_\delta(t)=\bmy(t/\delta^{2H};\bx)=\bx(t/\delta^{2H};\bx)-\vv
t/\delta^{2H}.
\]
as $\delta\to 0$. It turns out that the limit of $\bmy_\delta(t)$ is a
super-diffusive fractional Brownian motion $B_\alpha(t)$. This should
be contrasted with a standard Brownian motion limit for the process
$\bmy(t/\delta^2)$ obtained in \cite{kp79} for rapidly decorrelating
fields $\bF(\vx)$. We recall that a Gaussian, continuous trajectory,
$\mathbb R^d$--valued process $(B_\al(t))_{t\ge0}$ is called a
fractional Brownian motion (FBM) with Hurst exponent $\al\in(0,1)$ and
the covariance matrix ${\bf D}$ if for any $t,h\ge0$ the (normal)
random vector $B_\al(t+h)-B_\al(t)$ has mean zero and the covariance
${\bf D}h^{2\al}$.  In particular, $\E(B_\alpha(t)\otimes B_\alpha(t))={\bf D} t^{2\alpha}$
and the tracer propagation is close to ballistic as $\alpha\uparrow 1$
and correlations persist at infinity. Note that for $\al=1/2$ the FBM
becomes the "usual" Brownian motion matching the result of \cite{kp79}
for rapidly decayng correlations.

Anomalous diffusive behavior of passive scalars has been studied
extensively in the physics literature -- we mention
\cite{CF96,CFKL95,GK95,Kraichnan,SS95,W04} without any attempt at
completeness. However, the majority of physical papers consider
time-dependent (typically white in time) random flows with a non-zero
diffusivity.  Super-diffusive limits for particles in a random flow
have been also obtained rigorously in various situations -- for
instance, in \cite{AM1,AM2,AM3,AM4} for the random shear layer flows,
in \cite{fk-aap-00} for time-dependent random flows, in \cite{kola1}
for a mean zero, random flow with a positive molecular diffusivity and in
\cite{BA-O,O} for multiscale periodic flows. The novel aspect of the
present paper is that the flow is time-independent and the molecular
diffusivity is equal to zero.

The paper is organized as follows. The aforementioned formal
derivation of the fractional Brownian motion limit for
$\alpha\in(1/2,1)$ is presented in Section \ref{final}. We describe in
Section \ref{sec:main-results} upper and lower bounds for the mean
square displacement of trajectories that show the super-diffusive
scaling for $\alpha\in(1/2,1)$ and diffusive behavior for
$\alpha<1/2$. The main results are
Theorems~\ref{main1}~and~\ref{main2} -- they are proved in Sections
\ref{sec:pf-thm1} and
\ref{sec:pf-thm2}. Some auxiliary bounds are proved in Section
\ref{sec:corrector}. Appendix \ref{sec:append} contains a brief
review of multiple stochastic integration.

{\bf Acknowledgment.} LR was supported in part by an Alfred P. Sloan fellowship.
This work has been also supported by ONR and NSF grant DMS-0604687. 

\section{A formal limit for the advection equation}
\label{final}


\subsection*{The advection equation}

We present here a formal computation that leads to a FBM
limit for the solution of the advection equation
\begin{equation}\label{adv-1}
\pdr{\phi_\delta}{t}+(\vv+\delta \bF(\bx))\cdot\nabla_\bx \phi_\delta=0,
~~\phi_\delta(0,\vx)=u_0(\vx).
\end{equation}
Here $\vv\in\Rm^d$ is a fixed mean flow with $|\vv|=1$ and $\bF(\vx)$ is
a spatially homogeneous Gaussian random field with the two-point
correlation tensor $[R_{ij}(\bx)]$ of the form (\ref{covariance}) and the
power spectrum as in (\ref{pow-energy}). We can write then
$\bF(\vx)=\int e^{i\vx\cdot\bk}\hat \bF(d\bk),$ where  the spectral measure $\hat F(d\bk)$
satisfies
\begin{equation}\label{ft-F}
\E\left[\hat F_n(d\bk) \hat F_m(d\bp)\right]=\hat R_{nm}(\bk)\delta(\bk+\bp)d\bk d\bp.
\end{equation}
We will assume in this section that the exponent $\alpha\in(1/2,1)$ as
this is the range where we expect the FBM
behavior for a rescaled process in the limit $\delta\to 0$. Let us set
$H=1/(2\alpha)$ and introduce the rescaled time $t'=\delta^{2H}t$ and
the moving frame $\vx'=\vx+\vv t$ as well as the new unknown function
$u_\delta(t',\vx')=\phi_\delta\left(\frac{t'}{\delta^{2H}},\vx'-\vv
\frac{t'}{\delta^{2H}}\right)$. Then (\ref{adv-1}) becomes in the new variables
\begin{equation}\label{adv-2}
\pdr{u_\delta}{t'}+\delta^{1-2H} \bF\left(\vx'-\vv\frac{t'}{\delta^{2H}}\right)
\cdot\nabla_{\vx'} u_\delta=0,~~u_\delta(0,\vx')=u_0(\vx').
\end{equation}
For convenience
we introduce $\bG(t,\vx)=-\bF(\vx-\vv t)$ and drop the primes, arriving at
\begin{equation}\label{adv-3}
\pdr{u_\delta}{t}-\delta^{1-2H} \bG\left(\frac{t}{\delta^{2H}},\vx\right)\cdot\nabla_\vx u_\delta=0,
~~u_\delta(0,\vx)=u_0(\vx).
\end{equation}
The main goal of this section is to argue formally that
$\E\left[u_\delta(t,\vx)\right]$ converges, as $\delta\to 0$, to
\begin{equation}\label{fracbm}
\bar u(t,\vx)=\E\left[u_0(\vx+B_{D}(t))\right].
\end{equation}
Here $B_D(t)$ is the FBM with exponent $\alpha$ and   covariance matrix ${\bf D}=[D_{ij}]$, where
\begin{equation}\label{covar-fbm}
D_{ij}=\frac{1}{2\al^2}\,\int\frac{e^{i\bk\cdot\vv}}{|\bk|^{2\alpha+d-2}}\Gamma_{ij}(\hat{\bk})d\bk.
\end{equation}
We will establish this limit by evaluating term-wise the
expectation of the formal infinite Duhamel expansion for the
solution of (\ref{adv-3}) and computing the limit of the main
term.  We will neither attempt to justify the expansion, nor try
to estimate the error produced by the terms in the expansion which
are formally of a smaller order. The main difference
with the corresponding calculation in the case when correlations
are rapidly decaying (the Brownian motion limit case) is that all
Feynman diagrams contribute to the limit and not only the
so-called ladder diagrams.

\subsection*{An infinite expansion for the solution}

We rewrite \eqref{adv-3} as an integral equation:
\begin{eqnarray}\label{duhamel}
&&u_\delta(t,\vx)=
 u_0(\vx)+\delta^{1-2H}\int\limits_0^t
 \bG\left(\frac{s_1}{\delta^{2H}},\vx\right)\cdot
 \nabla_\bx u_\delta(s_1,\vx)ds_1.
\end{eqnarray}
Iterating (\ref{duhamel}) by substituting for $u_\delta(s_1,\cdot)$ appearing on the right side
we obtain a formal expansion
\begin{equation}\label{sum-N}
 u_\delta(t,\vx)=
\sum\limits_{n=0}^{+\infty}\delta^{(1-2H)n}\int\limits_{\Delta_n(t)}
 \bG^{(n)}\left(\frac{s_1}{\delta^{2H}},\ldots,
    \frac{s_n}{\delta^{2H}},\vx\right)ds_1\ldots ds_n
\end{equation}
Here, we have set $\bG^{(0)}\left(\bx\right):=u_0(\bx)$ and, assuming
that $\bG^{(n)}\left(s_1,\ldots,s_n,\bx\right)$ has been already defined for
some $n\ge1$, we let
\[
\bG^{(n+1)}\left(s_1,\ldots,
    s_n,s_{n+1},\bx\right):=\bG(s_{n+1},\bx)\cdot\nabla_\bx \bG^{(n)}\left(s_1,\ldots,
    s_n,\bx\right).
 \]
The time integration region appearing in (\ref{sum-N}) is the simplex
\[
\Delta_n(t):=[(s_1,\ldots,s_n)\in\R^n:0\le s_n\le\ldots\le s_1\le t].
\]
Passing to the Fourier transform in the definition of $G^{(n)}$ and using
induction we arrive at an explicit expression
\begin{eqnarray}
\label{G-n}
\bG^{(n)}\!\left(s_1,\ldots,s_n,\bx\right)=
i^n
\!\!\int\! e^{-i\vv\cdot(\bk_1s_1+\ldots+\bk_ns_n)}
e^{i\bx\cdot(\bk_0+\ldots+\bk_n)}
\prod\limits_{p=1}^n\left[
\left(\sum\limits_{l=0}^{p-1}\bk_{l}\right)\cdot
\hat{\bF}(d\bk_p)\right]
\hat u_0(\bk_0)d\bk_0.
\end{eqnarray}
Next, we take formally the expectation of the infinite series in (\ref{sum-N})
term-wise and obtain
\begin{equation}\label{exp-sum-N}
   \E [u_\delta(t,\vx)]=
\sum\limits_{n=0}^{+\infty}I_n(\delta)(t,\vx),
\end{equation}
where
\[
I_n(\delta)(t,\vx)=\delta^{2(1-2H)n}\int\limits_{\Delta_{2n}(t)}
\E\bG^{(2n)}\left(\frac{s_1}{\delta^{2H}},\ldots,
 \frac{s_{2n}}{\delta^{2H}},\vx\right)\prod\limits_{i=1}^{2n}
 ds_i.
\]
We have used above the fact that the expectation of a product of an
odd number of centered Gaussian variables equals zero. The next step is to evaluate
each of the terms $I_n(\delta)$.

\subsection*{The main term in the expansion}

We will compute the individual terms $I_n(\delta)$ using  Feynman diagrams.  Recall that given Gaussian random variables
$Y_1,\dots,Y_{2n}$ the expectation $\E\left(Y_1Y_2\dots
Y_{2n}\right)$ is
\begin{equation}\label{rule-gauss}
\E\left(Y_1Y_2\dots Y_{2n}\right)=
\sum_{{\cal F}\in{\mathfrak F(n)}}\prod_{\widehat{pq}\in{\cal F}}\E(Y_pY_q).
\end{equation}
Here $\mathfrak F(n)$ is the set of all permutations (complete Feynman diagrams) of
the elements of $\{1,\ldots,2n\}$  such that ${\cal F}\circ{\cal F}=Id$, and
${\cal F}(k)\neq k$ for all $1\le k\le 2n$. The notation
$\widehat{pq}\in{\cal F}$ means that ${\cal F}(p)=q$ (or that $p$ and $q$ are
connected by an edge in the Feynman diagram ${\cal F}$) and $p<q$.

Using (\ref{rule-gauss})  in the expression for $I_n(\delta)$ we obtain
from \eqref{G-n} and (\ref{exp-sum-N}) that
\begin{eqnarray*}
&&I_n(\delta)(t,\vx)=
(-1)^n\delta^{2(1-2H)n}\sum\limits_{{\cal F}\in\mathfrak F(n)}
\sum_{i_1,\ldots,i_{2n}=1}^d\mathop{\int\!\!\ldots\!\!\int}\limits_{\Delta_{2n}(t)}
e^{-i\vv\cdot(\bk_1s_1\delta^{-2H}+\ldots+\bk_{2n}s_{2n}\delta^{-2H})}\\
&&\times
 e^{i\bx\cdot(\bk_0+\bk_1+\ldots+\bk_{2n})}
  \prod\limits_{\widehat{pq}\in{\cal F}}\left\{\left(\sum\limits_{l=0}^{p-1}k_{l,i_p}\right)
\left(\sum\limits_{j=0}^{q-1}k_{j,i_q}\right)\E[\hat{F}_{i_p}(d\bk_p)\hat{F}_{i_q}(d\bk_q)]\right\}
\hat u_0(\bk_0)d\bk_0
\prod\limits_{i=1}^{2n} ds_i.
\end{eqnarray*}
After evaluating the expectation above, using (\ref{ft-F}), this expression becomes
\begin{eqnarray}\label{ind-1}
&&\!\!\!\!\! I_n(\delta)(t,\vx)=(-1)^n\delta^{2(1-2H)n}\sum\limits_{i_1,\ldots,i_{2n}=1}^{d}
\sum\limits_{{\cal F}\in\mathfrak F(n)}\mathop{\int\!\!\ldots\!\!\int}\limits_{\Delta_{2n}(t)}
\prod\limits_{\widehat{pq}\in {\cal F}}\Big\{\exp\Big\{i\vv\cdot\bk_p\frac{s_q-s_p}{\delta^{2H}}
 \Big\}
\\
&&\!\!\!\!\!
\times\Big(k_{0,i_p}+\sum\limits_{l=1}^{p-1}k_{l,i_p}\Big)
\Big(k_{0,i_q}+\sum\limits_{j=1}^{q-1}k_{j,i_q}\Big)
\hat R_{i_p i_q}(\bk_p)\delta(\bk_p+\bk_q)d\bk_pd\bk_q\Big\}e^{i\bk_0\cdot\vx}\hat u_0(\bk_0)
d\bk_0\prod\limits_{i=1}^{2n} ds_i.\nonumber
\end{eqnarray}
The leading order term in (\ref{ind-1}) is
\begin{eqnarray}
&&I_n^0(\delta)(t,\vx)=(-1)^n\delta^{2(1-2H)n}\sum\limits_{i_1,\ldots,i_{2n}=1}^{d}
\sum\limits_{{\cal
F}\in\mathfrak F(n)}\mathop{\int\!\!\ldots\!\!\int}\limits_{\Delta_{2n}(t)}\prod\limits_{\widehat{pq}
  \in {\cal F}}\left[\exp\left\{i
   \vv\cdot\bk_p\frac{s_q-s_p}{\delta^{2H}}
      \right\}\right.
\nonumber\\
&&
\left.\times k_{0,i_p}k_{0,i_q}
\hat R_{i_p i_q}(\bk_p)d\bk_p\right]e^{i\bk_0\cdot\vx}\hat u(\bk_0)
d\bk_0\prod\limits_{i=1}^{2n} ds_i\label{ind-23}\\
&&=(-1)^n\delta^{2(1-2H)n}
\sum\limits_{{\cal
F}\in\mathfrak F(n)}\mathop{\int\!\!\ldots\!\!\int}\limits_{\Delta_{2n}(t)}\prod\limits_{\widehat{pq}
  \in {\cal F}}\left[
R\left(\frac{s_q-s_p}{\delta^{2H}}\vv\right)\bk_0\cdot\bk_0
\right]e^{i\bk_0\cdot\vx}\hat u_0(\bk_0)
d\bk_0\prod\limits_{i=1}^{2n} ds_i.\nonumber
\end{eqnarray}
The reason why $I_n^0(\delta)(t,\vx)$ is indeed the main contribution to $I_n(\delta)(t,\vx)$ is as follows.
Note that the two-point correlation function $R_{ij}(\vx)$ decays algebraically for large $|\vx|$:
\begin{equation}\label{R-slarge}
R_{i j}\left(\vv s \right)=D_{ij}(s)s^{2\al-2},
\end{equation}
with the matrix $D_{ij}(s)$, which converges, as $s\to +\infty$, to
\begin{equation}\label{barD}
\bar D_{ij}=\int\frac{1}{|\bk|^{2\alpha+d-2}}\Gamma_{ij}(\hat{\bk})e^{i\bk\cdot\vv}d\bk
\end{equation}
(a more explicit expression for  $\bar D_{ij}$
using  Legendre polynomials may be obtained using the Hecke-Funk
theorem (see e.g. \cite{hochstadt}, p. 181) but we will not need it here).
Using the above information in (\ref{ind-23}) we obtain:
\[
I_n^0(\delta)(t,\vx)=(-1)^n\delta^{2(1-2H)n}
\!\!\!\!\sum\limits_{{\cal
F}\in\mathfrak F(n)}\mathop{\int\!\!\ldots\!\!\int}\limits_{\Delta_{2n}(t)} \!e^{i\bk_0\cdot\vx}
(\bar \bD\bk_0\cdot\bk_0)^n\hat u(\bk_0)
d\bk_0 \!\!\!\!\!\prod\limits_{\widehat{pq}
  \in {\cal F}}\!\!\left|\frac{s_q-s_p}{\delta^{2H}}\right|^{2\alpha-2}
\prod\limits_{i=1}^{2n} ds_i+o(1).
\]
where $\bar \bD=[\bar D_{ij}]$.
However, and this is the crucial point in this calculation, as $H=1/(2\alpha)$ we see that
\[
2(1-2H)n-2Hn(2\alpha-2)=0
\]
and the powers of $\delta$ exactly cancel each other.
We conclude that $I_n^0(\delta)(t,\vx)=J_n(t,\vx)+o(1)$, with
\[
J_n(t,\vx)=(-1)^n
\sum\limits_{{\cal
F}\in\mathfrak F(n)}\mathop{\int\!\!\ldots\!\!\int}\limits_{\Delta_{2n}(t)} e^{i\bk_0\cdot\vx}
(\bar \bD\bk_0\cdot\bk_0)^n\prod\limits_{\widehat{pq}
  \in {\cal F}}\left|{s_q-s_p}\right|^{2\alpha-2}
\hat u_0(\bk_0)
d\bk_0 \prod\limits_{i=1}^{2n} ds_i.
\]
In particular, $I_n^0(\delta)(t,\vx)$ is of order $O(1)$, as $\delta\to 0$. On
the other hand, all the other terms in (\ref{ind-1}) lead to expressions
similar to (\ref{ind-23}) but with $R_{ij}(\vx)$ replaced by its (possibly
higher order) derivatives with respect to some of the spatial
variables. However, derivatives of $R_{ij}(\vx)$ decay faster than
$|\vx|^{2-2\alpha}$ for large $|\vx|$ -- hence these terms produce a
too high power of $\delta$ as a factor, and vanish (at least,
term-wise) in the limit $\delta\to 0$.


\subsection*{Interpretation in terms of the fractional Brownian motion}

It remains now to relate $J_n(t,\bx)$ to the fractional Brownian motion and
sum all these terms. Note that the function
\[
f(s_1,\ldots,s_{2n}):=\sum\limits_{{\cal F}\in\mathfrak
F(n)}\prod\limits_{\widehat{pq}\in {\cal F}} \left|s_p-s_q\right|^{2\al-2}
\]
is symmetric in all of its arguments, that is,
$f(s_1,\ldots,s_{2n})=f(s_{\pi(1)},\ldots,s_{\pi(2n)})$, where
$\pi$ is an arbitrary permutation of $\{1,2.\dots,n\}$. Using this fact we
can rewrite $J_n(t,\vx)$  in the form
\begin{equation}\label{J-n-1}
J_n(t,\vx)=\frac{(-1)^n}{(2n)!}
\sum\limits_{{\cal F}\in\mathfrak F(n)}\int_0^t\ldots\int_0^t
    \prod\limits_{\widehat{pq}\in {\cal F}} \left|
    s_p-s_q
      \right|^{2\al-2}\prod\limits_{i=1}^{2n} ds_i
  \int e^{i\bk_0\cdot\vx}(\bar \bD \bk_0,\bk_0)^n\hat u_0(\bk_0)d\bk_0.
\end{equation}
A simple but useful observation is that
\[
\sum\limits_{{\cal F}\in\mathfrak F(n)}\prod\limits_{\widehat{pq}\in {\cal F}} \left|
    s_p-s_q
      \right|^{2\al-2}= c_\alpha^{2n}
\E\left[ \prod\limits_{p=1}^{2n} \int_{-\infty}^\infty\frac{\ie^{ik_ps_p}}{|k_p|^{\al-1/2}}w(dk_p)
   \right],
\]
where $w(dk_1),\ldots,w(dk_{2n})$ are independent Gaussian white
noises and $c_\al>0$ is given by
$$
c_\alpha:=\left(\frac{\Gamma(2\al-1)\sin(\pi\al)}{\pi}\right)^{1/2}.
$$
 This follows from the fact that
\[
\E\left[ \int_{-\infty}^\infty\frac{\ie^{ik_1s}}{|k_1|^{\al-1/2}}w(dk_1)
\int_{-\infty}^\infty\frac{\ie^{ik_2r}}{|k_2|^{\al-1/2}}w(dk_2)\right]=
\int _{-\infty}^\infty\frac{\ie^{ik_1(s-r)}}{|k_1|^{2\al-1}}dk_1=c_\alpha^{-2}|s-r|^{2\alpha-2}
\]
for
\[
\int_{-\infty}^\infty \frac{e^{ik}}{|k|^{2\alpha-1}}dk=2\int_0^{+\infty}
\frac{\cos k}{k^{2\alpha-1}}dk=\frac{\pi}{\Gamma(2\al-1)\sin(\pi\al)}.
\]
The last equality follows e.g. from  3), [539] of \cite{fichtenholz}.
Hence, $J_n(t,\vx)$ has a representation
\[
J_n(t,\vx)=\frac{(-1)^n}{(2n)!}
\int_0^t\ldots\int_0^t
   \E\left[ \prod\limits_{p=1}^{2n} \int\frac{\ie^{ik_ps_p}}{|k_p|^{\al-1/2}}w(dk_p)
   \right]\prod\limits_{i=1}^{2n} ds_i
  \int  e^{i\bk_0\cdot\vx}|c_\alpha|^{2n}(\bar \bD \bk_0,\bk_0)^n\hat u_0(\bk_0)d\bk_0.
\]
Performing now the integrations with respect to $s_i$ on the right side
we conclude that
\[
J_n(t,\vx)=
\frac{(-1)^n}{(2n)!}
   \E\left[ \prod\limits_{p=1}^{2n} \int\frac{\ie^{ik_pt}-1}{ik_p|k_p|^{\al-1/2}}w(dk_p)
   \right]
  \int|c_\alpha|^{2n} (\bar \bD \bk_0,\bk_0)^n\hat u_0(\bk_0)e^{i\bk_0\cdot\vx}d\bk_0.
\]
Next, using the harmonizable representation of the standard fractional Brownian motion,
see  Proposition 7.2.8, p. 328 of \cite{sataq},  we obtain
\[
J_n(t,\vx)=
\frac{(-1)^n}{(2n)!}c^{2n}|c_\alpha|^{2n} \E B_{\al}(t)^{2n}
  \int e^{i\bk_0\cdot\vx}(\bar \bD \bk_0,\bk_0)^n\hat u_0(\bk_0)d\bk_0.
\]
Here $B_\al(t)$ is a  fractional Brownian motion with the Hurst exponent $\al$
and
\[
c:=\left(\frac{\pi}{\al\Gamma(2\al)\sin(\al \pi)}\right)^{1/2}=
\left(\frac{\pi}{2\al^2\Gamma(2\al-1)\sin(\al \pi)}\right)^{1/2}
\]
Setting
$b:=cc_\alpha=1/(\alpha\sqrt{2})$
we may now re-write $J_n$ as
\begin{eqnarray*}
&&J_n(t,\vx)=
\frac{(-1)^n(2n-1)!!}{(2n)!}
t^{2n\al} b^{2n}\int e^{i\bk_0\cdot\vx}(\bar \bD \bk_0,\bk_0)^n\hat u_0(\bk_0)d\bk_0\\
&&~~~~~~~~~~
=\frac{1}{n!}
\int e^{i\bk_0\cdot\vx}\left[-\frac{b^2t^{2\alpha}}{2}(\bar \bD \bk_0,\bk_0)\right]^n\hat u_0(\bk_0)d\bk_0.
\end{eqnarray*}
Coming back to \eqref{exp-sum-N} we see that, as $\delta\to0$,
\begin{eqnarray*}
&&\E  u_\delta(t,\vx)\to\bar u(t,\vx)=\sum_{n=0}^\infty J_n(t,\vx)=
\int\exp\left\{i\bk_0\cdot\vx-\frac{b^2t^{2\alpha}}{2}(\bar \bD \bk_0,\bk_0)\right\}\hat u_0(\bk_0)d\bk_0\\
&&~~~~~~~~~~~~~~~~~~~~~~~~~~~~=
\int \exp\left\{i(\vx+\bB_{\bD}(t))\cdot\bk_0\right\}\hat u_0(\bk_0)d\bk_0=\E\left[u_0(\vx+\bB_{\bD}(t))\right],
\end{eqnarray*}
where $\bB_{\bD}(t)$ is a $d$--dimensional fractional Brownian motion
with the exponent $\alpha$ and the covariance matrix $\bD=\bar \bD/(2\al^2)$.
Therefore, we have formally established that (\ref{fracbm}) gives the
limit of $\E u_\delta(t,\vx)$.

In terms of the characteristics
\begin{equation}\label{basic-1}
     \frac{d\bX(t;\bx)}{dt}=\vv+\delta\bF(\bX(t;\bx)),\quad \bX(0;\bx)=\bx,
\end{equation}
for the original advection problem (\ref{adv-1}) we have formally argued that the one dimensional statistics
of the process
\[
\bmy(t;\vx)=\bX\left(\frac{t}{\delta^{2H}};\vx\right)-\frac{\vv t}{\delta^{2H}},
\]
with $H=1/(2\alpha)$, converge, as $\delta\to 0$, to the corresponding statistics of a fractional
Brownian motion with the exponent $\alpha$ and diffusion matrix $\bD$
given by (\ref{covar-fbm}).  In the next section we will show
rigorously that indeed the process $\bmy(t)$ behaves trivially 
(either vanishes or $\E[|\bmy(t)|^2]$ tends to infinity) on all time scales but
$t\sim O(\delta^{-2H})$.


\commentout{
{\bf Proof of Lemma \ref{lmrij}.}
To obtain \eqref{d-ij} note that
\begin{equation}\label{012401}
R_{i j}\left(\vv
    s
      \right)=-s^{2\al-2}\int\limits_0^{sK}\frac{dk}{k^{2\al-1}}\int\limits_{\mathbb S^{d-1}}\ie^{ik\vv\cdot\hat\bk}
      \hat k_i\hat k_jS(d\hat\bk).
\end{equation}
Since for $i\not=j$ the expression $ k_ik_j$ is a harmonic polynomial by virtue of the Hecke-Funk
theorem, see \cite{hochstadt} p. 181, we obtain that the right hand side of \eqref{012401} equals
\begin{equation}\label{012402}
-s^{2\al-2}v_iv_j\int\limits_0^{sK}\frac{dk}{k^{2\al-1}}\int\limits_{-1}^1\ie^{i k t}P_{2,d}(t)
(1-t^2)^{(d-3)/2}dt,
\end{equation}
which implies \eqref{d-ij}.
As for $i=j$ we
write
\begin{equation}\label{012403}
R_{i i}\left(\vv
    s
      \right)=-s^{2\al-2}\int\limits_0^{sK}\frac{dk}{k^{2\al-1}}\int\limits_{\mathbb S^{d-1}}\ie^{ik\vv\cdot\hat\bk}
      \left(\frac{|\hat\bk|^2}{d}-\hat k_i^2\right)S(d\hat\bk)
\end{equation}
$$
-s^{2\al-2}\left(1-\frac1d\right)
\int\limits_0^{sK}\frac{dk}{k^{2\al-1}}\int\limits_{\mathbb S^{d-1}}\ie^{ik\vv\cdot\hat\bk}
     S(d\hat\bk).
$$
The right hand side of \eqref{012403} can be again calculated with the help of the Hecke-Funk
theorem but this time with
 $(k_1^2+\ldots+k_d^2)/d-k_1^2$ and $\bone$ as the respective choices of the harmonic polynomials.
 To get \eqref{d-ii} one uses the fact that $P_{0,d}(t)=\bone$.
$\Box$
}

\section{The main results: rigorous bounds for the trajectories}
\label{sec:main-results}

\subsection*{Preliminaries}\label{sec2.1}

{\bf The probability space.} Suppose that  $m$ is a positive integer  and
$\vartheta_\rho(\bx):=(1+|\bx|^2)^{-\rho}$, $\bx\in\R^d$, where
$\rho>d/2$. Let $\Om$ be the Hilbert space of $d$-dimensional
incompressible vector fields that is the completion of the space
$C_{0,div}^{\infty}:=\{\om\in C_0^\infty(\mathbb R^d;\mathbb
R^d):\nabla_\bx\cdot \om=0\}$ with respect to the norm
\[
\|\om\|^2_{\Om}:=\int\limits_{\mathbb
R^d}(|\om(\bx)|^2+|\nabla_{\bx} \om (\bx)|^2+\cdots
+|\nabla^m_{\bx}\om(\bx)|^2)\vartheta_{\rho}(\bx)\,d\bx.
\]
 We
shall assume that $m>d/2+1$ so that any $\om\in\Om$ is of $C^1$ class
of regularity by the Sobolev embedding theorem.

\noindent{\bf The random field.} The random field  is
set to be simply $\bF(\bx;\om):=\omega(\vx)$.
Denote also $\bF(\om)=(F_1(\om),\ldots,F_d(\om)):=\om(\bze)$.
We suppose that  $\P$ is a Borel measure given on   $\Om$  that satisfies the following
hypotheses:

(H1) it is \emph{Gaussian}, that is, for any $N\ge1$, $\bx_1,\ldots,\bx_N\in\mathbb R^d$
we have $\bF(\bx_1),\ldots,\bF(\bx_N)$ is a Gaussian, $Nd$--dimensional random vector,

(H2) it is \emph{centered} and \emph{homogeneous}, that is, $\bF$ is of mean zero
and the two-point correlation matrix  depends only on the relative position of the points:
\[
\int F_i(\bz;\omega)F_j(\by;\omega)d\P(\omega)=R_{ij}(\bz-\by),
\]
where $\bR(\bx)=[R_{ij}(\bx)]$ is given by \eqref{covariance}. As the measure $\P$ is Gaussian, this
condition guarantees that $\P$ is invariant with respect to any spatial shift transformation
$\tau_\vx:\Om\to\Om$, $\bx\in\R^d$
defined by $\tau_{\bx}\om(\bz):=\om(\bx+\bz)$.
The existence of such a measure on $\Omega$ is guaranteed e.g.  by the results of \cite{fkp},
see Section 2.3.
Thanks to the assumed form of the power spectrum we may suppose
that the realizations of the velocity field are $\P$--a.s. analytic in the $\bx$ variable, see e.g.
\cite{be}. We will denote by $\E$ the mathematical expectation with respect to the measure $\P$.

\subsection*{The main results}
We consider the particle trajectory
given as the solution to \eqref{basic-1} with the starting point $\bx=\bze$ and 
with the random field $\bF(\vx)$ constructed in the previous section,
and define the particle deviation from the mean position
$
\by(t)=\bX(t;\bze)-\vv t.
$
We also introduce
\begin{equation}\label{Y-capital}
Y_{i}(t):=\int_0^t \E[y_i(s)]^2ds,\quad i=1,\ldots,d.
\end{equation}
Suppose that the times $T_\delta>0$ are such that
$\lim_{\delta\to0+}T_\delta=+\infty$. Define
the Cesaro limit of the mean square of the
fluctuation amplitude as
\begin{equation}\label{cesaro}
C\hbox{-}\!\!\lim\limits_{\delta\to0+}\E\left|\by\left(T_\delta\right)\right|^2
   :=\lim\limits_{\delta\to0+}\,\frac{1}{T_\delta}\,Y\left(T_\delta\right),
\end{equation}
provided that the limit on the right hand side exists, whether it is finite, or not.

We will distinguish two cases: as we have mentioned in the Introduction,
when $\alpha\in(1/2,1)$ the Kubo-Taylor formula (\ref{kubo-form}) diverges and we expect
a behavior different from the usual diffusive limit. On the other hand, when $\alpha<1/2$
the diffusion coefficient given by (\ref{kubo-form}) remains finite and the usual diffusive
behavior would not be surprising.

\subsubsection*{The case $\al\in(1/2,1)$}

We have argued formally in Section \ref{final} that in this range the process $\by(t)$ converges on the time scale
$t\sim O(\delta^{-2H})$, $H=1/(2\alpha)$ to a fractional Brownian
motion with the exponent $\alpha$. Our first result confirms the predicted time-scale
in this range of the parameter $\alpha$.
\begin{thm}
\label{main1} Suppose that $t>0$, $\al\in(1/2,1)$ and
$\rho>0$. Let  $T_\delta^+:=\delta^{-2H(1+\rho)}t^{1+\rho}$
and
$T_\delta^-:=\delta^{-2H(1-\rho)}t^{1-\rho}$. Under the  above assumptions about
the random field  $\bF(\bx)$  we have
\begin{equation}\label{140408}
C\hbox{-}\!\!\lim\limits_{\delta\to0+}\E\left|\by\left(T_\delta^+\right)\right|^2
=+\infty
\end{equation}
and
\begin{equation}\label{140409}
   \lim\limits_{\delta\to0+}\E\left|\by\left(T_\delta^-\right)\right|^2=0.
\end{equation}
\end{thm}
\subsubsection*{The case $\al<1/2$}
In this case we expect that $\by(t)$ behaves diffusively on the time
scale $t\sim O(\delta^{-2})$, as in the situation when the correlation
tensor decays rapidly in space, that is, when $\alpha$ is very
negative. This time scale is confirmed by the next theorem.
\begin{thm}
\label{main2} Suppose that $\al<1/2$. Then, for arbitrary $t,\rho>0$ we have
\begin{equation}\label{102606}
C\hbox{-}\!\!\lim\limits_{\delta\to0+}\delta^{-\rho}
\E\left|\by\left(t\delta^{-2}\right)\right|^2=+\infty
\end{equation}
and
\begin{equation}\label{112606}
\lim\limits_{\delta\to0+}\E\left|\by\left(t\delta^{-2(1-\rho)}\right)\right|^2=0.
\end{equation}
\end{thm}
Theorems~\ref{main1} and~\ref{main2} are proved in
Sections~\ref{sec:pf-thm1} and~\ref{sec:pf-thm2}, respectively.

\section{The proof of Theorem \ref{main1}}\label{sec:pf-thm1}

\subsection*{The lower bound}

We prove the lower bound (\ref{140408}). This is done
with the help of a general lower bound, which relates the long time
behavior of the trajectory to the behavior of the resolvent near
the border of the spectrum $\lambda=0$.  In order to formulate it let us begin with some
preliminary definitions.

{\bf The basic spaces.} For
any $p\in[1,+\infty)$ and $\phi\in L^p(\P)$ we adopt the notation $D_k \phi:=\frac{d}{d h}
\phi(\tau_{h\bbe_k}\om)|_{h=0}$, where $\bbe_k$, $k=1,\cdots,d$ is the
$k$-th vector of the canonical basis in $\mathbb R^d$. The derivatives are
understood in the $L^p$ sense.
Let $
W^{p,m}$ be the Banach space consisting of those elements $\phi\in
L^p(\P)$, for which
\[
\|\phi\|_{p,m}^p:=\sum\limits_{i_1+\cdots+i_d\leq
m}\|D_{1}^{i_1}\cdots D_d^{i_d} \phi\|_{L^p}^p<+\infty.
\]
We set
\begin{equation}\label{C}
{\cal C}:=\bigcap_{1<p<+\infty}
W^{p,2}.
\end{equation}

{\bf The spectral measure.} The spectral theorem (see e.g. Theorem 1.4.2, p. 18 of \cite{rozanov}) implies
 that there
exists a complex vector valued spectral measure
$\hat{\vF}(\cdot)=(\hat{F}_1(\cdot),\ldots,\hat{F}_d(\cdot))$
 defined
over $(\R^d,{\cal B}(\R^d))$ whose components take values in
$L^2(\P)$ such that
\begin{equation}
\bF(\bx)=\int \mbox{e}^{i\bx\cdot\bk}\hat{\vF}(d\bk). \label{a.0}
\end{equation}
 The spectral measure is Gaussian, that is, for any set $A\in{\cal B}(\R^d)$
 the $2d$--dimensional random vector $(\mbox{Re }\!\hat{\vF}(A),\mbox{Im }\!\hat{\vF}(A))$
is Gaussian. Since the random field \eqref{a.0} is real vector valued
we must have $\hat{\vF}^*(d\bk)=\hat{\vF}(-d\bk)$.
The corresponding structure measure equals
  \begin{equation}\label{structure}
\E[\hat{F}_i(d\bk)\hat{F}_j^*(d\bk')]=
\hat R_{ij}(\bk)
\delta(\bk-\bk')d\bk\,d\bk',\quad i,j=1,\ldots,d.
\end{equation}

{\bf The Hermite polynomials.} Suppose that $\psi=(\psi_1,\ldots,\psi_d)$, where
$\psi_i\in C_0^\infty(\R^d)$, $i=1,\dots,d$ are  complex valued, even functions, that is,
$\psi_i(-\bk)=\psi^*_i(\bk)$. We
write
\begin{equation}\label{060616}
\int\psi(\bk)\cdot\hat{\bF}(d\bk):=\sum_{i=1}^d\int\psi_i(\bk)\hat{F}_i(d\bk).
\end{equation}
Denote by $H$ the subspace of $L^2(\P)$ obtained by taking the closure
of the linear span of the elements of the form \eqref{060616}. It is a Gaussian
Hilbert space in the sense of Definition 1.2 p. 4 of \cite{janson}.
We can define then, see Definition 2.1 of ibid., the space of the $n$-th
degree polynomials ${\cal P}_n$ as the $L^2$--closure of
all the elements of the form $p(\phi_1,\ldots,\phi_k)$,
where $p(\cdot)$ is  polynomial of at most $n$-th degree
with real valued coefficients and $\phi_i\in H$, $i=1,\ldots,k$.
It can easily be seen that ${\cal P}_n$ can be also characterized as
the $L^2$--closure of the space spanned by
\begin{equation}\label{10615}
\int\ldots\int \psi(\bk_1,\ldots,\bk_n)\cdot \hat{\bF}(d\bk_1)\otimes\ldots\otimes\hat{\bF}(d\bk_n),
\end{equation}
where $\psi:(\mathbb R^d)^n\to  (\mathbb C^d)^n$ is even, that is,
$\psi(-\bk_1,\ldots,-\bk_n)=\psi^*(\bk_1,\ldots,\bk_n)$, -- see
the Appendix  for the definition of the multiple stochastic integral.
For any $n\ge0$ we denote by $H^{:n:}:={\cal P}_n\ominus{\cal P}_{n-1}$ \emph{the space of the
$n$--th degree Hermite polynomials}, see Definition 2.1 of \cite{janson}.
Here ${\cal P}_{-1}:=\{0\}$.
It is well known, see Theorem 2.6 of ibid., that $L^2(\P)=\bigoplus_{n\ge0} H^{:n:}$

\subsubsection*{A variational principle for the resolvent}\label{sec:resolvent}

We define the random field $\vV(\bx;\om):=\vv+\delta\bF(\bx;\om)$ and
let $\bX(t)$ be the solution of \eqref{basic} with the initial condition $\bx$ set at $\bze$.
 The {\it environment process} is 
 given by $\eta(t):=\tau_{\bX(t)}\om$.
 It is an $\Om$--valued, deterministic, dynamical system,
 with $\P$ as its invariant measure. The corresponding group
 of Koopman operators $P^tf(\om)=f\eta_t(\om),$ $t\ge0$  extends then to a $C_0$--continuous,
 unitary group on $L^2(\P)$.
Its generator $L$
is given by
\[
L\phi=(\vv+\delta\bF)\cdot\nabla \phi,\,\,\mbox{ for }\phi\in{\cal
C}.
\]
Here $\nabla \phi:=(D_1\phi,\ldots,D_d\phi)$. One can show, in the same way as it was done in see Lemma 4.1 of \cite{kola},
 that ${\cal C}$ is invariant under the action of the group $(P^t)$ and,
 since it is dense in $L^2(\P)$, it is a core of the generator.

Denote by $R_\la\phi:=(\la-L)^{-1}\phi$ the resolvent operator
defined for any $\la>0$ and $\phi\in L^2(\P)$.
 We can formulate now the variational principle
that will be crucial for us in the sequel, cf. Lemma 2.1 of \cite{bernardin}.
\begin{prop}
\label{prop10021} For any $f\in L^2(\P)$ and $\la>0$ we have
\begin{equation}\label{62105}
  \langle R_\la f,f\rangle_{L^2(\P)}=
  \sup\left[2\langle f,\phi\rangle_{L^2(\P)}-\frac{1}{\la}\|L\phi\|_{L^2(\P)}^2-
\la\|\phi\|_{L^2(\P)}^2:\,\phi\in{\cal
  C}\right].
  \end{equation}
\end{prop}
{\bf Proof.}
Let $R_\la^{s}:=(\lambda-L)^{-1}_s$  be the symmetric part of the (bounded) operator  $R_\la$
and let ${\cal R}$ be the range of $R_\la^{s}$.
As $L$ is anti-selfadjoint,  $R_\la^{s}$ is given explicitly by
\begin{equation}\label{rlas}
R_\la^sf=\frac{1}{2}\left[(\lambda-L)^{-1}+(\lambda+L)^{-1}\right]f
=\lambda(\lambda-L)^{-1}(\lambda+L)^{-1}f,
~~\hbox{for any $f\in L^2(\P)$.}
\end{equation}
An elementary calculation shows that for any $f\in L^2(\P)$ we have, as ${\cal R}\in D(L)$,
\begin{eqnarray}\label{021703}
&& \langle R_\la f,f\rangle_{L^2(\P)}=\langle R_\la^{s} f,f\rangle_{L^2(\P)}
=\sup[2\langle
f,\phi\rangle_{L^2(\P)}-
\langle \phi,(R_\la^{s})^{-1}\phi\rangle_{L^2(\P)}:\,\phi\in {\cal R}]
\\
&&~~~~~~~~~~~~~~=\sup\left[2\langle
f,\phi\rangle_{L^2(\P)}-\frac{1}{\la}
\langle (\lambda-L)\phi,(\lambda-L)\phi\rangle_{L^2(\P)}:\,\phi\in {\cal R}\right]\nonumber\\
&&~~~~~~~~~~~~~~=\sup\left[2\langle
f,\phi\rangle_{L^2(\P)}-\la\|\phi\|_{L^2(\P)}^2-\frac{1}{\la}\|L\phi\|^2_{L^2(\P)}
:\,\phi\in {\cal R}\right].\nonumber
\end{eqnarray}
The last equality above uses anti-selfadjointness of $L$.  As ${\cal
C}$ is a core of $L$ we can use it instead of $\cal R$ as the set of
test functions in the variational principles \eqref{021703} -- thus,
(\ref{62105}) follows. $\Box$

\subsubsection*{A lower bound on the variance of trajectory fluctuations}
\label{sec2.3}

The variational principle (\ref{62105}) for the resolvent is used as
follows.  Let $\bG=(G_1,\ldots,G_d)\in L^2(\mathbb P)$, the next
result concerns the lower bound of the second absolute moment of
$\bmz(t):=\int_0^t\bG(\eta(s))ds$ (here $\eta(t)$ is the environment process), cf. Lemma 2, p. 655 of
\cite{kola1}. To formulate it we introduce
\begin{equation}\label{140402}
Z_{i}(t):=\int_0^t \E[z_i(s)]^2ds=2\int_0^t\left[\int\limits_0^s\left(\int\limits_0^{s_1}
\langle P^{s_2}G_i,G_i\rangle_{L^2(\P)}ds_2\right)ds_1\right]ds
\end{equation}
and $Z(t)=\sum_{i=1}^dZ_{i}(t)$.
Note that from \eqref{140402} we obtain, in particular that
\begin{equation}\label{140403}
    Z_{i}(t)\le t^3\|G_i\|_{L^2(\P)}^2/3.
\end{equation}
The following proposition relates the small $\lambda$ behavior of the resolvent to the large time behavior
of the process $Z(t)$.
 \begin{prop}
 \label{prop130402}
Let $\lambda>0$ and define
\begin{equation}\label{180405}
G(\la):=\sum\limits_{i=1}^d\langle R_{\la}G_i,G_i\rangle_{L^2(\P)}.
\end{equation}
Assume that
  $t,\bt,\rho>0$ are given. Let $T_\delta:=t\delta^{-\beta}$.
Then, there exist  constants $C_*,\delta_0>0$
 such that
\begin{equation}\label{62104b}
Z\left(T_\delta^{1+\rho}\right)\ge \,
C_*\,T_\delta^2\,
G(T_\delta^{-1}),\quad\forall\,\delta\in(0,\delta_0].
\end{equation}
 \end{prop}
 {\bf Proof.}
Using \eqref{140402} and integration by parts we find that
\begin{equation}\label{140401}
\int\limits_0^{+\infty}\ie^{-\la\,t}Z_{i}(t)\,dt
=2\mathop{\int\!\int\!\int\!\int}\limits_{0\le s_2\le s_1\le s\le t}
e^{-\lambda t}\langle P^{s_2}G_i,G_i\rangle_{L^2(\P)}ds_1ds_2dsdt
\end{equation}
$$
=2\lambda^{-3}\int\limits_0^{+\infty}e^{-\lambda s_2}\langle P^{s_2}G_i,G_i\rangle_{L^2(\P)}ds_2
=
2\lambda^{-3}\langle R_{\la}G_i,G_i\rangle_{L^2(\P)}.
$$
On the other hand, since $t\mapsto Z_{i}(t)$ is an increasing function, we can write that
the utmost left hand side of \eqref{140401} is bounded from above as
\begin{eqnarray}\label{140405}
&&\int\limits_0^{+\infty}\ie^{-\la\,t}Z_{i}(t)\,dt
\le
Z_{i}\left(\la^{-(1+\rho)}\right)\int\limits_0^{\la^{-(1+\rho)}}\ie^{-\la t}
\,dt+
\int\limits_{\la^{-(1+\rho)}}^{+\infty}\ie^{-\la t}
Z_{i}\left(t\right)\,dt.
\\
&&
\!\!{\le}\la^{-1}Z_{i}\left(\la^{-(1+\rho)}\right)
 +
\frac{1}{3}\|G_i\|_{L^2(\P)}^2\int\limits_{\la^{-(1+\rho)}}^{+\infty}\ie^{-\la t}
t^3\,dt\nonumber
\le \la^{-1}Z_{i}\left(\la^{-(1+\rho)}\right)+\frac{C\|G_i\|_{L^2}^2}{3\la^3}
\int\limits_{\la^{-(1+\rho)}}^{+\infty}\ie^{-(\la t)/2}dt\nonumber
\end{eqnarray}
for some absolute constant $C>0$. We have used (\ref{140403}) in the second inequality above.
Performing now the integration
on the utmost right hand side of \eqref{140405} and recalling \eqref{140401} we obtain that
\begin{equation}\label{140406}
2\lambda^{-3}\langle R_{\la}G_i,G_i\rangle_{L^2(\P)}\le
\la^{-1}Z_i\left(\la^{-(1+\rho)}\right)+\frac{C\ie^{-\la^{-\rho} }\|G_i\|_{L^2(\P)}^2}{3\la^4}.
\end{equation}
Summing up over $i$ we
obtain \eqref{62104b} upon choosing $\la :=T_\delta^{-1}$.
$\Box$

\subsubsection*{The proof of the lower bound}

The lower bound (\ref{140408}) in Theorem \ref{main1} is proved as follows: note that
\[
\by(t)=\delta\int_0^t \bF(\eta(s))ds
\]
is of the form of the functionals considered in Proposition
\ref{prop130402} (with $\bG=\delta \bF$). Therefore, to prove
(\ref{140408}) we will first use suitable test functions in the
variational principle (\ref{62105}) for the resolvent, and then use  (\ref{62104b}) to obtain
the lower bound for $Y(t)$ given by (\ref{Y-capital}).

The test functions are chosen as follows: suppose that $\psi(\bk)\in
C_0^\infty(\R^d)$ is a smooth complex valued, even vector function:
$\psi_i(-\bk)=\psi_i^*(\bk)$, $i=1,\dots,d$ and denote by ${\cal P}_1$ the family of
all random elements of the form
\begin{equation}\label{190403}
\phi(\omega)=\int\psi(\bk)\cdot\hat{\bf F}(d\bk;\omega).
\end{equation}
By virtue of \eqref{62105} we have a lower bound
\begin{equation}\label{140410}
  \langle R_\la F_i,F_i\rangle_{L^2}\ge
  \sup\left[2\langle F_i,\phi\rangle_{L^2}-\frac{1}{\la}\|L\phi\|_{L^2}^2-
\la\|\phi\|_{L^2}^2:\,\phi\in{\cal P}_1\right].
  \end{equation}
Observe that
\begin{equation}\label{90303}
  2\langle
F_i,\phi\rangle_{L^2}=2\int\limits_{\mathbb R^d}
 \,\psi(\bk)\cdot\Gamma(\hat\bk){\bf e}_i\,\frac{a(|\bk|)d\bk}{|\bk|^{2\alpha+d-2}},
\end{equation}
where $\Gamma(\hat\bk)=[\Gamma_{ij}(\hat\bk)]$.
Since we have
\begin{equation}\label{030123}
    D_j \phi={i}\, \int\limits_{\mathbb R^d}\,k_j\,\psi(\bk)\cdot\hat{\bF}(d\bk),
\end{equation}
the operator $L$ acts as
\begin{equation}\label{190405}
L\phi=(\vv+\delta\bF)\cdot\nabla
\phi={i}\left[~\int\limits_{\mathbb R^d}\,\vv\cdot\bk\,\,\psi(\bk)\cdot\hat{\bF}(d\bk)+
\delta\int\limits_{\mathbb R^d}\int\limits_{\mathbb R^d}\psi(\bk)
\cdot\hat{\bF}(d\bk)\,\bk\cdot\hat{\bF}(d\bk')\right]
\end{equation}
and
\begin{equation}\label{90304}
  \|\phi\|_{L^2(\P)}^2= \int\limits_{\mathbb R^d}\,
  [\Gamma(\hat\bk)\psi(\bk)\cdot\psi(\bk)]\,\frac{a(|\bk|)d\bk}{|\bk|^{2\alpha+d-2}}
 \end{equation}
we have
\begin{eqnarray}\label{180401}
&&\|L\phi\|_{L^2(\P)}^2=
\int\limits_{\mathbb R^d}
\frac{(\vv\cdot\bk)^2[\Gamma(\hat\bk)\psi(\bk)\cdot\psi(\bk)]a(|\bk|)d\bk}{|\bk|^{2\al+d-2}}\\
&&+\delta^2\int\limits_{\mathbb R^d}\int\limits_{\mathbb R^d}
\frac{[\Gamma(\hat\bk)\psi(\bk)\cdot\psi(\bk)]
[\Gamma(\hat\bk')\bk\cdot\bk ]a(|\bk|)a(|\bk'|)d\bk d\bk'}{(|\bk||\bk'|)^{2\al+d-2}}
\nonumber\\
&&+\delta^2\int\limits_{\mathbb R^d}
\int\limits_{\mathbb R^d}\frac{[\Gamma(\hat\bk)\psi(\bk)\cdot\bk'][\Gamma(\hat\bk')\bk\cdot\psi(\bk')]
a(|\bk|)a(|\bk'|)d\bk d\bk'}{
 (|\bk||\bk'|)^{2\al+d-2}}.
\nonumber
\end{eqnarray}
Using an elementary estimate
\begin{equation}\label{190407}
(\Gamma(\hat\bk)\psi(\bk)\cdot\bk')\le
(\Gamma(\hat\bk)\psi(\bk)\cdot\psi(\bk))^{1/2}(\Gamma(\hat\bk)\bk'\cdot\bk')^{1/2}
\end{equation}
we conclude that
\begin{equation}\label{180403}
 \langle R_\la F_i,F_i\rangle_{L^2}\ge \sup[{\mathfrak J}_i(\phi):\,\phi\in{\cal P}_1],
\end{equation}
where ${\mathfrak J}_i(\phi)$ is a quadratic functional given by
\begin{eqnarray}\label{180402}
&&    {\mathfrak J}_i(\phi):=2\int\limits_{\mathbb R^d}
 \,[\psi(\bk)\cdot\Gamma(\hat\bk){\bf e}_i]\,\frac{a(|\bk|)d\bk}{|\bk|^{2\alpha+d-2}}
-\lambda\int\limits_{\mathbb R^d}\,
  [\Gamma(\bk)\psi(\bk)\cdot\psi(\bk)]\,\frac{a(|\bk|)d\bk}{|\bk|^{2\alpha+d-2}}\\
&& \!\!\!\!- \frac1\lambda\int\limits_{\mathbb R^d}
\frac{(\vv\cdot\bk)^2[\Gamma(\hat\bk)\psi(\bk)\cdot\psi(\bk)]a(|\bk|)d\bk}{|\bk|^{2\al+d-2}}
-\frac{2\delta^2}{\lambda}\int\limits_{\mathbb R^{2d}}
 \frac{[\Gamma(\hat\bk)\psi(\bk)\cdot\psi(\bk)][\Gamma(\hat\bk')\bk\cdot\bk]
a(|\bk|)a(|\bk'|)d\bk d\bk'}{(|\bk||\bk'|)^{2\al+d-2}}.
\nonumber
\end{eqnarray}
The maximizer of the functional given above equals
\begin{equation}\label{maxim}
    \psi_*(\bk)=\Gamma(\hat\bk){\bf e}_i\left\{\lambda+
    \frac{1}{\la}\left[(\vv\cdot\bk)^2+\delta^2|\bk|^2{\cal H}\right]\right\}^{-1},
\end{equation}
where
\[
{\cal H}:=
2\int\limits_{\mathbb R^d}\Gamma(\hat\bk'){\bf e}_1\cdot{\bf e}_1\,
    \frac{ \,a(|\bk'|) d\bk'}{|\bk'|^{2\al+d-2}}=
2\left(1-\frac{1}{d}\right)\int\limits_{\mathbb R^d}\,
    \frac{ \,a(|\bk'|) d\bk'}{|\bk'|^{2\al+d-2}}>0.
\]
Thus, from \eqref{180403} we obtain
\begin{equation}\label{180403b}
P(\la):=\sum\limits_{i=1}^d \langle R_\la F_i,F_i\rangle_{L^2}\ge m(\delta),
\end{equation}
where
\begin{equation}\label{m-delta}
m(\delta):=
(d-1)\int\limits_{\mathbb R^d}\left\{\lambda+
    \frac{1}{\la}\left[(\vv\cdot\bk)^2+\delta^2|\bk|^2{\cal H}\right]\right\}^{-1}
    \,\frac{a(|\bk|)d\bk}{|\bk|^{2\alpha+d-2}}.
\end{equation}

We claim that $m(\delta)\ge C\lambda^{1-2\alpha}$ for $0<\lambda\le 1$.
To show this we will assume with no loss of generality that $\vv={\bf e}_1$.
The expression in \eqref{m-delta}
is of the same order of magnitude as
\begin{equation}\label{180407}
\int\limits_{|\bk|\le 1}\frac{\la d\bk}{(\lambda^2+
  k_1^2+2\delta^2|\bk|^2{\cal H})|\bk|^{2\alpha+d-2}}
\ge C\la\int\limits_{|\bk|\le 1}
    \,\frac{d\bk}{(\lambda^2+
  k_1^2+\delta^2|\bk|^2)|\bk|^{2\alpha+d-2}}
\end{equation}
for some $C>0$.
Writing $\bk=(k_1,\bml)$ we can further transform the
right hand side of \eqref{180407}. It equals
\begin{equation}\label{180408}
C\la\int\limits_{k_1^2+l^2\le 1}
    \,\frac{l^{d-2}dk_1dl}{[\lambda^2+
  (1+\delta^2)k_1^2+\delta^2l^2](k_1^2+l^2)^{\alpha+d/2-1}},
\end{equation}
where $l=|\bml|$. Introducing the polar coordinates
$l=\varrho\cos\theta$, $k_1=\varrho\sin\theta$
we can estimate \eqref{180408} from below by
\begin{equation}\label{180409}
C\la\int\limits_{0}^{1}\int\limits_{0}^{\pi/2}
    \,\frac{\cos^{d-2}\theta d\varrho d\theta}{(\lambda^2+
  \varrho^2\sin^2\theta+\varrho^2\delta^2)\varrho^{2\alpha-1}}\ge
  C_1\la\int\limits_{0}^{1}\int\limits_{0}^{\pi/4}
    \,\frac{ d\varrho d\theta}{[\lambda^2+
  \varrho^2(\theta^2+\delta^2)]\varrho^{2\alpha-1}}.
\end{equation}
Substituting $\varrho':=\la^{-1}\varrho\sqrt{\theta^2+\delta^2}$ we
see that the right hand side of \eqref{180409} equals
\[
 C_1\la^{1-2\al}\int\limits_{0}^{\pi/4}\frac{d\theta}{(\theta^2+\delta^2)^{1-\al}}
 \left[\int\limits_{0}^{(\theta^2+\delta^2)^{1/2}\la^{-1}}
    \,\frac{ d\varrho}{(1+
  \varrho^2)\varrho^{2\alpha-1}}\right].
\]
Since $2-2\al<1$, the first integral above converges even for
$\delta=0$, and the second also has a finite limit as $\lambda\to 0$.
We conclude from the above and (\ref{180403b}) that
\begin{equation}\label{180410}
    P(\la)\ge C_2\la^{1-2\al},\quad\,\forall\,\la\in(0,1]
\end{equation}
and some $C_2>0$.

With the lower bound (\ref{180410}) at hand we are ready to
finish the proof of Theorem \ref{main1}.
 Let $\rho'>0$ be arbitrary and let
$T_\delta^+$ be as in the statement of the theorem. We
apply Proposition \ref{prop130402} with
$T_\delta:=(T_\delta^+)^{1/(1+\rho')}$.
 Then, for sufficiently small
$\delta_0$ we have
\[
Y\left(T_\delta^+\right)=Y\left(T_\delta^{1+\rho'}\right)
\ge \,
C_3\,\delta^2T_\delta^2P(T_\delta^{-1})
\stackrel{\eqref{180410}}{\ge}C_4\delta^2T_\delta^{2\al+1}
\]
for all $\delta\in(0,\delta_0]$ and some $C_3,C_4>0$. Hence,
\[
\frac{1}{T_\delta^+}\,Y\left(T_\delta^+\right)\ge
C_4\delta^2T_\delta^{2\al-\rho'}=C_4\delta^{2-2H(1+\rho)(2\al-\rho')/(1+\rho')}
t^{(1+\rho)(2\al-\rho')/(1+\rho')}
\]
and \eqref{140408} follows, provided that $\rho'>0$ is chosen sufficiently small.

\subsection*{The upper bound}

The upper bound (\ref{140409}) in Theorem \ref{main1} is proved using an approximation
of the trajectory by the correctors.

\subsubsection*{The corrector fields}
\label{sec2.4}

Let  $\bF^{(1)}_{\la}:=\bF$ and for any $\la>0$ we define the corrector
field of the first order in the direction of $\vec e_p$
\begin{equation}\label{030501}
    \chi_{p,\la}^{(1)}:=\int\frac{1}{\la-\mbox{i}\vv\cdot\bk}\,\hat{F}_p(d\bk),\quad p=1,\ldots,d.
\end{equation}
Note that $\chi_{p,\la}^{(1)}$ satisfies
\[
(\la-\vv\cdot\nabla)\chi_{p,\la}^{(1)}=F^{(1)}_{p,\la},\quad p=1,\ldots,d.
\]
Let us define
$\bF^{(2)}_{\la}=(F^{(2)}_{1,\la},\ldots, F^{(2)}_{d,\la})$, where
\begin{equation}\label{030502}
F^{(2)}_{p,\la}:=\bF\cdot\nabla\chi_{p,\la}^{(1)}
={i}\int\!\!\int\frac{\bk_1\cdot\hat{\bF}(d\bk_2)}{\la-\mbox{i}\vv\cdot\bk_1}\,\hat{F}_p(d\bk_1).
\end{equation}
Then, we may write, using the definition of the first order corrector
\begin{equation}\label{030503}
x_p(t)-v_p t=\delta \int\limits_0^tF_p(\eta(s))ds=
    \delta\la\int\limits_0^t\chi_{p,\la}^{(1)}(\eta(s))ds-
    \delta\int\limits_0^t\vv\cdot\nabla\chi_{p,\la}^{(1)}(\eta(s))ds.
\end{equation}
On the other hand, we also have
\begin{eqnarray}\label{030504}
&& \chi_{p,\la}^{(1)}(\eta(t))-\chi_{p,\la}^{(1)}(\om)=
    \int\limits_0^t (\vv+\delta\bF(\eta(s)))\cdot\nabla\chi_{p,\la}^{(1)}(\eta(s))ds
   \\
&&
=
    \int\limits_0^t \vv\cdot\nabla\chi_{p,\la}^{(1)}(\eta(s))ds+
    \delta\int\limits_0^t \bF^{(2)}(\eta(s))\cdot\nabla\chi_{p,\la}^{(1)}(\eta(s))ds.\nonumber
\end{eqnarray}
so that (\ref{030503}) becomes
\begin{equation}\label{030505}
    x_p(t)-v_p t=\delta^2\int\limits_0^t F_{p,\la}^{(2)}(\eta(s))ds
+\delta\la\int\limits_0^t\chi_{p,\la}^{(1)}(\eta(s))ds+ \delta\chi_{p,\la}^{(1)}(\om)
-\delta  \chi_{p,\la}^{(1)}(\eta(t))
 \end{equation}
Now, we may iteratively define $F_{p,\la}^{(n)}:=\bF\cdot\nabla\chi_{\la,p}^{(n-1)}$ and
let $\chi_{\la,p}^{(n)}$ be the solution of
\begin{equation}\label{031506}
    (\la-\vv\cdot\nabla)\chi_{\la,p}^{(n)}=F_{p,\la}^{(n)}.
\end{equation}
Then, for any $n\ge1$ we have a decomposition
\begin{equation}\label{030505b}
    x_p(t)-v_p t=\delta^{n}\int\limits_0^t F_{p,\la}^{(n)}(\eta(s))ds
+\la\sum\limits_{l=1}^{n-1}\delta^l\int\limits_0^t\chi_{p,\la}^{(l)}(\eta(s))ds
+ \sum\limits_{l=1}^{n-1}\delta^l[\chi_{p,\la}^{(l)}(\om)
-  \chi_{p,\la}^{(l)}(\eta(t))].
\end{equation}
Technically,  the most important  results of this section are the following bounds for the correctors.
\begin{prop}
\label{corr-est}
Suppose that $\al\in(1/2,1)$. Then, for each $n\ge1$ we have
\begin{equation}\label{051506}
    \|\chi_{\la}^{(n)}\|_{L^2}\le \frac{C}{\la^{n\al}}\left(1+\left|\log \la\right|\right)^{n/2},\quad\la\in(0,1]
\end{equation}
for some constant $C>0$. When, on the other hand $\al<1/2$ we have
\begin{equation}\label{071506}
    \|\chi_{\la}^{(n)}\|_{L^2}\le \frac{C}{\la^{n/2}}\left(1+\left|\log \la\right|\right)^{n/2},\quad\la\in(0,1].
\end{equation}
\end{prop}
As a consequence we obtain the following.
\begin{cor}
\label{cor-F}
For some constant $C>0$ we have
\begin{equation}\label{051506b}
    \|\bF_{\la}^{(n)}\|_{L^2}\le \frac{C}{\la^{n\al}}\left(1+\left|\log \la\right|\right)^{n/2},\quad\la\in(0,1]
\end{equation}
for $\al\in(1/2,1)$ and
\begin{equation}\label{071506b}
    \|\bF_{\la}^{(n)}\|_{L^2}\le \frac{C}{\la^{n/2}}\left(1+\left|\log \la\right|\right)^{n/2},\quad\la\in(0,1]
\end{equation}
for $\al<1/2$.
\end{cor}
The proof of Proposition \ref{corr-est} is rather technical and we postpone it until Section \ref{sec:corrector},
where the proof of Corollary \ref{cor-F} can also be found.

In what follows we shall also need the following lemma which shows that the inner products of correctors with
$\bF$ are of a smaller order than one would naively expect from the Cauchy-Schwartz inequality and
Proposition \ref{corr-est}.
\begin{lemma}
\label{scal-prod}
Suppose that $\al<1/2$. Then,
for a given $n\ge1$ there exists a constant $C>0$ such that
\begin{equation}\label{sc-pr}
|\langle
\chi_{\la,p}^{(n)},F_p\rangle_{L^2}|\le \frac{C}{\la^{(n-1)/2}}\left(1+|\log\la|\right)^{(n+1)/2}
,\quad\forall\,\la\in(0,1].
\end{equation}\end{lemma}
This lemma is also proved in Section \ref{sec:corrector}.


\subsubsection*{The proof of the upper bound}

%

We now prove \eqref{140409}.
Using \eqref{030505b} we estimate
\begin{equation}\label{022606}
    \E\left|\by\left(T_\delta^-\right)\right|^2\le
    C\left[\delta^{2n}(T_\delta^-)^2\|\bF_{\la}^{(n)}\|_{L^2}^2\right.
\left. +\sum\limits_{l=1}^{n-1}\delta^{2l}(\la T_\delta^-)^2\sum_{p=1}^d\|\chi_{p,\la}^{(l)}\|_{L^2}^2
+\sum\limits_{l=1}^{n-1}\delta^{2l}\sum_{p=1}^d\|\chi_{p,\la}^{(l)}\|_{L^2}^2\right]
\end{equation}
Recall that
$T_\delta^-=\delta^{-2H(1-\rho)}t^{1-\rho}$. This together with estimates
\eqref{051506} and \eqref{051506b} imply that the right hand side of \eqref{022606}
can be estimated by
$$
C\left[(\delta\la^{-\al})^{2n}(T_\delta^-)^2
+\sum\limits_{l=1}^{n-1}(\delta\la^{-\al})^{2l}(\la T_\delta^-)^2
+\sum\limits_{l=1}^{n-1}(\delta\la^{-\al})^{2l}\right].
$$
Choose $\la:=\delta^{2H(1-\rho)}$, we obtain then
\begin{equation}\label{032606}
    \E\left|\by\left(T_\delta^-\right)\right|^2\le
    Ct^{2(1-\rho)}\left(\delta^{n\rho}\delta^{-4H(1-\rho)}
+\sum\limits_{l=1}^{n-1}\delta^{l\rho}\right).
 \end{equation}
For an arbitrary $\rho>0$ we can choose $n$ sufficiently large so that the right hand side
of \eqref{032606} is of order of magnitude $o(1)$ so that \eqref{140409} follows.
$\Box$

\section{The proof of Theorem \ref{main2}}
\label{sec:pf-thm2}

For the most part the proof of the upper bound (\ref{112606}) in
Theorem \ref{main2} is a repetition of what has been done in the
corresponding situation in the previous case. Observe that
\eqref{022606} still holds
 with $T_\delta^-:=t\delta^{-2(1-\rho)}$.
We can choose now $\la:=(T_\delta^-)^{-1}$ and easily convince
ourselves that \eqref{112606} holds.

The heuristic reason why this direction is simpler once the upper
bound in Theorem \ref{main1} has been obtained is that the time
$\delta^{-2}$ is much shorter that $\delta^{-2H}$, $H=1/(2\alpha)$ for
$\alpha<1/2$. Therefore the fact that ``nothing happens until the time
$O(\delta^{-2})$'' is not very surprising in Theorem \ref{main2}. The lower bound in Theorem \ref{main2}
is more informative -- it tells that something happens at the time $O(\delta^{-2})$ which is much earlier
than the time scale $\delta^{-2H}$.

We now prove the lower bound (\ref{102606}).
Let $\rho'>0$  and $T_\delta:=(t\delta^{-2})^{1/(1+\rho')}$. We shall further specify
the parameter $\rho'$  later on. By virtue of Proposition \ref{prop130402} we conclude that
for a certain $C>0$ and sufficiently small $\delta>0$
\begin{equation}\label{142706}
\,Y\left(T_\delta^{1+\rho'}\right)\ge C \delta^2T_\delta^{2}P(T_\delta^{-1}),
\end{equation}
where $P(\la)$ is given by
\eqref{180403b}.
Recall that the correctors $\chi_{\la,p}^{(n)}$ are the solutions of \eqref{031506}
for an arbitrary $\la>0$ and $N\ge n\ge1$. Let the remainders
$r_{\la,p}^{(N)}$ be given by
\begin{equation}\label{122706}
    [\la-(\vv+\delta\bF)\cdot\nabla]r_{\la,p}^{(N)}=F_{p,\la}^{(N+1)}.
\end{equation}
Multiplying both sides of \eqref{122706} by $r_{\la,p}^{(N)}$ and integrating out
we get an obvious bound
\begin{equation}\label{132706}
    \|r_{\la,p}^{(N)}\|_{L^2}\le \frac{1}{\la}\|F_{p,\la}^{(N+1)}\|_{L^2}\stackrel{\eqref{071506b}}
    {\le} \frac{C}{\la^{1+(N+1)/2}}(1+|\log\la|)^{N/2}.
\end{equation}
Observe also that
\[
R_\la F_p=\sum\limits_{n=1}^N\delta^{n-1}\chi_{\la,p}^{(n)}+\delta^Nr_{\la,p}^{(N)}
\]
and therefore
\[
\langle R_\la F_p,F_p\rangle_{L^2}\ge\langle
\chi_{\la,p}^{(1)},F_p\rangle_{L^2}-C\sum\limits_{n=1}^{N-1}\delta^{n}|\langle
\chi_{\la,p}^{(n+1)},F_p\rangle_{L^2}|-C\la^{-3/2}(\delta\la^{-1/2})^{N}(1+|\log\la|)^{N/2}.
 \]
Lemma \ref{scal-prod} allows us to obtain the following estimate
\begin{equation}\label{152706}
\langle R_\la F_p,F_p\rangle_{L^2}\ge\langle
\chi_{\la,p}^{(1)},F_p\rangle_{L^2}
-C\sum\limits_{n=1}^{N-1}(\delta\la^{-1/2})^{n}|(1+|\log\la|)^{n/2}
-C\la^{-3/2}(\delta\la^{-1/2})^{N}(1+|\log\la|)^{N/2}.
\end{equation}
Choosing $\la:=T_{\delta}^{-1}$ we obtain from \eqref{142706} and
\eqref{152706}, with sufficiently large $N$ chosen,
 that
\begin{equation}\label{162706}
\,\delta^{2-\rho}Y\left(t\delta^{-2}\right)\ge C t^{2/(1+\rho')} \delta^{-\rho+4\rho'/(1+\rho')},
\end{equation}
which clearly implies \eqref{102606}, provided that $-\rho+4\rho'/(1+\rho')<0$.
$\Box$

\section{The proofs of the corrector bounds}\label{sec:corrector}

In this section we prove the technical bounds on the correctors
stated in Proposition~\ref{corr-est}, Corollary~\ref{cor-F} and
Lemma~\ref{scal-prod}.

\subsubsection*{The Feynman diagrams}

The proofs of the corrector bounds make an extensive use of the
Feynman diagrams. Let us recall now the corresponding basic notions.
Let $Z_n:=\{1,\ldots,n\}$, a {\em Feynman diagram} $\cF$ (of order
$n\geq 0$ and rank $r\geq0$) based on $Z_n$ is a graph consisting of a
set $B(\cF)\subset Z_n$ of vertices from $Z_n$, and a set $E(\cF)$ of
$e(\cF)$ edges connecting points in $Z_n$ without common endpoints. So
there are $e(\cF)$ pairs of vertices, each joined by an edge, and
$a(\cF):=n-2e(\cF)$ unpaired vertices, called {\em free vertices}. An
edge whose endpoints are $m,n\in B$ is denoted by $\widehat{mn}$
(unless otherwise specified, we always assume $m<n$). A diagram
${\cal F}$ is said to be {\em
 based on} $B(\cF)$. Denote the set of
free vertices by $A({\cal
 F})$, so $A({\cal F})=\cF\setminus E(\cF)$.
The diagram is {\em
 complete} if $A({\cal F})$ is empty and {\em
incomplete},  otherwise. Also for a given $1\le p\le n$ let us denote
by $A_p(\cF)$ the set of all free vertices that are less or equal to $p$ and by
$L_p(\cF)$ the union of $A_p(\cF)$ and all the left vertices of edges
$\widehat{mn}$ for which $m\le p<n$. Let $L(\cF):=L_n(\cF)$  denote
the left vertices of all edges belonging to $E(\cF)$.

\subsection*{Expressions for correctors}

We begin with some explicit expressions for the correctors
$\chi_{m,\lambda}^{(n)}$.  Recall that the functions $\chi_{m,\lambda}^{(n)}$, $m=1,2\dots,d$,  are
defined iteratively as the solutions of
\begin{equation}\label{corr-n-eq-2}
(\la-\vv\cdot\nabla)\chi_{m,\la}^{(n)}=F_{m,\la}^{(n)},~~
F_{m,\la}^{(n)}:=\bF\cdot\nabla\chi_{m,\lambda}^{(n-1)},
\end{equation}
with $\chi_{m,\lambda}^{(1)}$, the solution of
\[
(\la-\vv\cdot\nabla)\chi_{m,\la}^{(1)}=F_{m,\la},\quad m=1,\ldots,d,
\]
given explicitly by
\begin{equation}\label{corr-1-eq-2}
\chi_{m,\la}^{(1)}:=\int\frac{1}{\la-\mbox{i}\vv\cdot\bk}\,\hat{F}_m(d\bk),\quad m=1,\ldots,d.
\end{equation}
Let us introduce auxiliary functions
 \[
 \tilde h_n(\bk_1,\ldots,\bk_n):=\frac{\sum_{l=1}^{n}\bk_l}{\la-\mbox{ i}\vv\cdot(\sum_{l=1}^{n}\bk_l)}
 \]
 for $n\ge1$ and
 \[
 h_1^{(m)}(\bk_1):=\frac{{\bf e}_m}{\la-\mbox{ i}\vv\cdot\bk_1},~~~~~
 h_n(\bk_1,\ldots,\bk_n):=
\frac{\sum_{l=1}^{n-1}\bk_l}{\la-\mbox{ i}\vv\cdot(\sum_{l=1}^{n}\bk_l)},\quad n\ge2.
\]
Then, a simple induction argument shows that
the $n$--th order corrector in the direction of the vector ${\bf e}_p$ is given by
\begin{equation}\label{011506}
\chi_{m,\la}^{(n)}:= i^{n-1}\int h_1^{(m)}(\bk_1)\,
\otimes\,h_2(\bk_1,\bk_2)\,\otimes\ldots
\otimes\,h_n(\bk_1,\ldots,\bk_n)\,
\cdot\hat{\bF}(d\bk_1)\,\otimes\,\hat{\bF}(d\bk_2)\ldots\otimes\,\hat{\bF}(d\bk_n)
\end{equation}
and the fields $\bF^{(n)}_{\la}$ defined in (\ref{corr-n-eq-2}) are
$\bF^{(n)}_{\la}=(F^{(n)}_{1,\la},\ldots, F^{(n)}_{d,\la})$, where
\begin{equation}\label{021506}
F_{m,\la}^{(n)}:={i}^{n-2}\int {\bf e}_m\otimes\tilde h_1(\bk_1)\,
\otimes\ldots \otimes\,\tilde h_{n-1}(\bk_1,\ldots,\bk_{n-1})
\cdot\hat{\bF}(d\bk_1)\,\otimes\,\hat{\bF}(d\bk_2)\ldots\otimes\,\hat{\bF}(d\bk_{n}).
\end{equation}

\subsection*{The proof of Proposition \ref{corr-est}}

\subsubsection*{The Feynman diagrams expansion}

We now prove Proposition \ref{corr-est}, an $L^2$-bound for the
correctors.  Using the Feynman diagram expansion \eqref{a21} of a
multiple stochastic integral from the Appendix we can write that
\begin{equation}\label{fey-exp}
\chi_{\la}^{(n)}=\sum\limits_{\cal F}\chi_{\la}^{(n)}({\cal F}),
\end{equation}
where the summation extends over all Feynman diagrams ${\cal F}$ of the set
 $Z_n$. For a given  diagram $\cF$ with
free vertices $A(\cF)=\{n_1,\ldots,n_a\}$ and edges $E(\cF)=\{\widehat{pq}\}$ we
set
\[
\chi_{m,\la}^{(n)}({\cal F}):=:\int\!\ldots\!
\int f^{m,\la}(\bk_{n_1},\ldots,\bk_{n_a})\cdot\hat{\bF}(d\bk_{n_1})
\,\otimes\ldots\otimes\,\hat{\bF}(d\bk_{n_a}):,
\]
where $:\cdot:$ denotes the orthogonal projection onto the space of $a$--th degree
Hermite polynomials $H^{:a:}$, cf Definition 3.1 of \cite{janson},
and $f^{m,\la}:(\Rm^d)^a\to (\Rm^d)^a$ is given by
\[
f^{m,\la}_{i_{n_1},\ldots,i_{n_a}}(\bk_{n_1},\ldots,\bk_{n_a}):=\!
\int h_{1,i_1}^{(m)}(\bk_1)\ldots
h_{n,i_n}(\bk_1,\ldots,\bk_n)\!\!\!\prod_{\widehat{pq}\in E(\cF)}\!\!\!
\Gamma_{i_p,i_q}(\hat\bk_p)\frac{a(|\bk_p|)\delta(\bk_p+\bk_q)}{|\bk_p|^{2\al+d-2}}d\bk_p d\bk_q.
\]
Using Theorem 3.9 p. 26 of \cite{janson} one concludes
that
\begin{eqnarray}
&&\|\chi_{m,\la}^{(n)}({\cal F})\|_{L^2}^2=
\sum\limits_{\pi}\int f^{m,\la}(\bk_{n_1},\ldots,\bk_{n_a})
f^{m,\la}(\bk_{\pi(n_1)}',\ldots,\bk_{\pi(n_a)}')
\prod_{p=1}^a
\Gamma_{n_p,n_{\pi(p)}}(\hat\bk_{n_p})\frac{a(|\bk_{n_p}|)}
{|\bk_{n_p}|^{2\al+d-2}}\nonumber\\
&&~~~~~~~~~~~~~~~~~\times\delta(\bk_{n_p}+\bk_{\pi(n_p)}')
d\bk_{n_p} d\bk_{\pi(n_p)}',\label{010616}
\end{eqnarray}
where the summation extends over all permutations $\pi:A(\cF)\to A(\cF)$.
We assume with no loss of generality that $\vv={\bf e}_1$.
Also we write $\bk_i=(k_i,\bml_i)\in\R\times\R^{d-1}$, splitting out the first component of the
vector $\bk_i$. Then, with this notation
we have
\begin{eqnarray*}
&&|f^{m,\la}_{i_{n_1},\dots,i_{n_a}}(\bk_{n_1},\ldots,\bk_{n_a})|\le
\int\limits_{-K}^K\!\dots\!\int\limits_{-K}^K
\prod_{j=1}^n\frac{1}{\left|\la-\mbox{ i}\sum\limits_{l=1}^{j}k_l\right|}
\prod_{\widehat{pq}\in E(\cF)}
\frac{\delta(k_p+k_q)}{|k_p|^{2\al-1}}dk_p dk_q
\\
&&~~~~~~~~~~~~~~~~~~~~~~~~~~~~~~~~~~~~
\times\left\{\mathop{\int\!\!\ldots\!\!\int}\limits_{|\bml_p|\le K} \prod_{\widehat{pq}\in E(\cF)}
\frac{\delta(\bml_p+\bml_q)d\bml_p d\bml_q}{(k_p^2+|\bml_p|^2)^{(d-1)/2}}\right\}.
\end{eqnarray*}
The last integral can be estimated by
$C\prod_{\widehat{pq}\in E(\cF)}(1+\log^+|k_p|)$ so we conclude that
the right hand side of \eqref{010616} can be estimated by
\begin{equation}\label{020616}
a(\cF)!\int |f^{m,\la}|^2(\bk_{n_1},\ldots,\bk_{n_a})
\prod_{p=1}^a
\frac{a(|\bk_{n_p}|)d\bk_{n_p}}{|\bk_{n_p}|^{2\al+d-2}}
\end{equation}
$$
\le C\!
\int\limits_{-K}^K\!\!\ldots\!\!\int\limits_{-K}^K\! \tilde f_{\la}^2(k_{n_1},\ldots,k_{n_a})\prod_{p=1}^a
\frac{(1+\log^+|k_{n_p}|)dk_{n_p}}{|k_{n_p}|^{2\al-1}},
$$
where
\[
\tilde f_{\la}(k_{n_1},\ldots,k_{n_a}):=
\int\limits_{-K}^K\!\ldots\!\int\limits_{-K}^K \prod_{j=1}^n
\frac{1}{\left|\la-\mbox{ i}\sum\limits_{l=1}^{j}k_l\right|}\prod_{\widehat{pq}\in E(\cF)}
\frac{(1+\log^+|k_p|)\delta(k_p+k_q)}{|k_p|^{2\al-1}}dk_p dk_q.
\]

\subsubsection*{The proof of (\ref{051506})}

Suppose first that $\al\in(1/2,1)$ -- we need to prove the estimate
(\ref{051506}) from Proposition~\ref{corr-est}. Note that after
changing variables $k_j=\lambda k_j'$ and dropping the primes we get
the estimate of the left hand side of \eqref{020616} as
\[
\|\chi_{m,\la}^{(n)}({\cal F})\|_{L^2}^2\le C\la^{-2\al n}(1+|\log\la|)^nI,
\]
where
\[
I:=
\int_{-\infty}^\infty\dots\int_{-\infty}^\infty g^2(k_{n_1},\ldots,k_{n_a})
\prod_{p=1}^a
\frac{(1+\log^+|k_{n_p}|)dk_{n_p}}{|k_{n_p}|^{2\al-1}},
\]
with
\[
g(k_{n_1},\ldots,k_{n_a}):=\int\limits_{-\infty}^\infty
\ldots\int\limits_{-\infty}^\infty\prod_{j=1}^n\frac{1}{1+|\sum\limits_{l=1}^{j}k_l|}
\prod_{\widehat{pq}\in E(\cF)}
\frac{(1+\log^+|k_{p}|)\delta(k_p+k_q)}{|k_p|^{2\al-1}}dk_p dk_q.
\]
We shall show by an elementary calculation that $I<+\infty$ -- this is all that remains to prove
the estimate (\ref{051506}). We present the details for the convenience of the reader: note that
\begin{eqnarray*}
&&I=
\mathop{\int\ldots\int}_{\Rm^{a+4e}}
\prod_{j=1}^n\frac{1}{1+|\sum_{l=1}^{j}k_l|}\times
\prod_{j'=1}^{n}\frac{1}{1+|\sum_{l'=1}^{j'}k_{l'}'|}
\times\prod_{m=1}^a
\frac{(1+\log^+|k_{n_m}|)dk_{n_m}}{|k_{n_m}|^{2\al-1}}\\
&&~~~~\times\prod_{\widehat{pq}\in E(\cF)}
\frac{(1+\log^+|k_{p}|)\delta(k_p+k_q)dk_p dk_q}{|k_p|^{2\al-1}}\times\prod_{\widehat{p'q'}\in E(\cF)}
\frac{(1+\log^+|k_{p'}|)\delta(k_{p'}+k_{q'})dk_{p'}'dk_{q'}}{|k_{p'}|^{2\al-1}}\\
&&~~~~=\int\limits_{\Rm^{n}}
\prod_{j=1}^n\frac{1}{1+|\sum_{l=1}^{j}k_l|}\times
\prod_{j'=1}^{n}\frac{1}{1+|\sum_{l'=1}^{j'}k_{l'}'|}
\times\prod_{m=1}^a
\frac{(1+\log^+|k_{n_m}|)dk_{n_m}}{|k_{n_m}|^{2\al-1}}\\
&&~~~~\times\prod_{p\in L(\cF)}
\frac{(1+\log^+|k_{p}|)dk_p }{|k_p|^{2\al-1}}\times\prod_{p'\in L(\cF)}
\frac{(1+\log^+|k_{p'}|)dk_{p'}'}{|k_{p'}|^{2\al-1}}.
\end{eqnarray*}
We adopt the rule that in the sums above
$k_r+k_s=0$ and $k_r'+k_s'=0$, if $\widehat{rs}$ is an edge.
Now, the integration over $|k|\le 1$ is not a problem so in order to show that $I<+\infty$ it
suffices only to prove that
\begin{eqnarray*}
&&I'=\int_{\Rm^{n}}
\prod_{j=1}^n\frac{1}{1+|\sum_{l=1}^{j}k_l|}\times
\prod_{j'=1}^{n}\frac{1}{1+|\sum_{l'=1}^{j'}k_{l'}'|}\\
&&
\times\prod_{m=1}^a
\frac{(1+\log^+|k_{n_m}|)dk_{n_m}}{1+|k_{n_m}|^{2\al-1}}\times\prod_{p\in L(\cF)}
\frac{(1+\log^+|k_{p}|)dk_p }{1+|k_p|^{2\al-1}}\times\prod_{p'\in L(\cF)}
\frac{(1+\log^+|k_{p'}|)dk_{p'}'}{1+|k_{p'}|^{2\al-1}}<+\infty.
\end{eqnarray*}
Using  H\"older inequality with $\bar p$ such that $\bar p(2\alpha-1)=2$ and $q=2/(3-2\alpha)>1$ such that
$1/\bar p+1/q=1$ leads to
$
I'\le C(II)^{1/q},$
with
\begin{eqnarray*}
&&II=\int\limits_{\Rm^{n}}\prod_{p=1}^n\frac{1}{1+|\sum_{l=1}^{p}k_l|^q}\times
\prod_{p'=1}^{n}\frac{1}{1+|\sum_{l'=1}^{p'}k_{l'}'|^q}
\times\prod_{{r}\in A(\cF)}
dk_r\times\prod_{{m}\in L(\cF)}
dk_m\times\prod_{{r'}\in L(\cF)}dk_{r'}'.
\end{eqnarray*}
Now, we introduce new variables $\eta_1,\dots,\eta_n$ as follows:
first, we look at the sums $S_m=\sum_{j=1}^mk_j$ with $m=1,\dots,n$,
with the terms $k_r+k_s=0$ if $\widehat{rs}$ is an edge.  We say that
$\eta_1=k_1$, then we take $j_2$ that is the smallest number larger
than one which is either free or a left vertex, and let
$\eta_2=S_{j_2}$. Note that either $\eta_2=k_2$ if $k_2$ is not a
right end, or $\eta_2=k_3$ if $(k_1k_2)$ is an edge. We continue in
the same way: having defined $\eta_{l-1}=S_{j_{l-1}}$ we take $j_l$ to
be the first index larger than $j_{l-1}$ which is not a right end of
an edge, and take $\eta_{l}=S_{j_{l}}$. In this way we will pick
$n-e({\cal F})=a({\cal F})+e({\cal F})$ sums out of $S_m$,
$m=1,\dots,n$ and will define $\eta_1,\eta_2,\dots,\eta_{n-e({\cal
F})}$. Note that the change of variables is lower-triangular. Then, we
do a similar procedure with the $k'$-variables -- we look at
$P_m=\sum_{j=1}^mk_j'$ and set $\eta_{n-e+l}=P_{q_l}$, $l=1,\ldots,e$,
where $q_1<\ldots<q_e$ are the left vertices.  This defines all
$\eta_l$, $l=1,\dots,n$ and keeps the change of variables lower-triangular.
Hence, it is invertible with the Jacobian equal to
$1$. As a consequence, we may drop all sums in the denominator which are
not equal to one of $\eta_j$ and we have
\begin{eqnarray*}
&&II\le \int\limits_{\Rm^{n}}
\prod_{j=1}^{n}\frac{1}{1+\left|\eta_j\right|^q}d\eta_j<+\infty.
\end{eqnarray*}
It follows that $I<+\infty$ and the proof of (\ref{051506}) is complete.
\subsubsection*{The proof of (\ref{071506})}

It remains to prove the estimate (\ref{071506}) in
Proposition~\ref{corr-est}.  Suppose now that $\al<1/2$.
 The right
hand side of \eqref{010616} can be estimated by
\begin{equation}\label{010626}
I:= C\int\limits_{-K}^K\ldots\int\limits_{-K}^K\tilde f_{\la}^2(k_{n_1},\ldots,k_{n_a})
\prod_{p=1}^a
dk_{n_p}
,
\end{equation}
where
\[
\tilde f_{\la}(k_{n_1},\ldots,k_{n_a}):=
\int\limits_{-K}^K\!\ldots\!\int\limits_{-K}^K \prod_{j=1}^n\frac{1}{\left|\la-{ i}\sum\limits_{l=1}^{j}k_l\right|}
\prod_{\widehat{pq}\in E(\cF)}
\delta(k_p+k_q)dk_p dk_q.
\]
We have
\begin{eqnarray*}
&&I\le C
\int_{-K}^{K}\dots\int_{-K}^{K}
\prod_{j=1}^n\frac{1}{\la+|\sum_{l=1}^{j}k_l|}\times
\prod_{j'=1}^{n}\frac{1}{\la+|\sum_{l'=1}^{j'}k_{l'}'|}\\
&&
\times\prod_{m=1}^a
dk_{n_m}\times\prod_{\widehat{pq}\in E(\cF)}
\delta(k_p+k_q)dk_p dk_q\times\prod_{\widehat{p'q'}\in E(\cF)}
\delta(k_{p'}+k_{q'})dk_{p'}'dk_{q'}.
\end{eqnarray*}
We introduce the new variables $\eta_1,\dots,\eta_n$ as in the
previous case.  With the help of this transformation and replacing the
remaining $n$ of the denominators above by $1/\lambda$ we conclude
that
\begin{equation}\label{010701}
\|\chi_{\la}^{(n)}({\cal F})\|_{L^2}^2\le \frac{C}{\la^{n}}\int_{-nK}^{nK}\dots\int_{-nK}^{nK}
\prod_{j=1}^{n}\frac{1}{\la+\left|\eta_j\right|}d\eta_j\le \frac{C}{\la^{n}}
\left[\log\left(\frac{1}{\la}\right)+1\right]^n
\end{equation}
for some $C>0$ and all $\la\in(0,1]$. Thus \eqref{071506} follows.
$\Box$

\subsection*{The proof of Corollary \ref{cor-F}}

We now prove the estimate (\ref{071506b}) in Corollary~\ref{cor-F}.
Using the Cauchy-Schwartz inequality we obtain
\[
 \|\bF_{\la}^{(n)}\|_{L^2}\le\|\bF\|_{L^4}\|\nabla\chi_{\la}^{(n)}\|_{L^4}.
\]
However, since all the $L^p$ norms, $1\le p<+\infty$ on ${\cal P}_n$
are equivalent, see Theorem 5.10 p. 62 of \cite{janson},
 and $\bF\in{\cal P}_1$, $\nabla\chi_{\la}^{(n)}\in{\cal P}_n$ we conclude that
\[
 \|\bF_{\la}^{(n)}\|_{L^2}\le C\|\bF\|_{L^2}\|\nabla\chi_{\la}^{(n)}\|_{L^2}
\]
for some constant $C>0$. Thanks to the fact that the spectral measure $\hat{\bF}$
has a compact support the gradient operator $\nabla$ restricted to ${\cal P}_n$ is  bounded
(this can be seen by a direct calculation $D_p$ on elements of the form \eqref{10615}).
This in turn implies \eqref{071506b}. $\Box$

\subsection*{The proof of Lemma \ref{scal-prod}}

Finally, we prove the estimate (\ref{sc-pr}) in Lemma \ref{scal-prod}.
We only need to be concerned with $n$ odd, for otherwise the
left hand side of \eqref{sc-pr} vanishes.
Using  expansion \eqref{fey-exp}
we can write that
\begin{equation}\label{020701}
\langle
\chi_{\la,p}^{(n)},F_p\rangle_{L^2}=\sum\limits_{\cal F}\langle
\chi_{\la,p}^{(n)}({\cal F}),F_p\rangle_{L^2}
\end{equation}
and the summation extends over those Feynman diagrams $\cal F$ for which $a(\cF)=1$.
Suppose that $q$ is
the only free vertex of $\cF$. Then,
\begin{eqnarray*}
&&\!\!\!\langle
\chi_{\la,p}^{(n)}({\cal F}),F_p\rangle_{L^2}=\sum_{i_1,\ldots,i_n}\int\!\ldots\!
\int h_{1,i_1}^{(p)}(\bk_1)\ldots
h_{n,i_n}(\bk_1,\ldots,\bk_n)\Gamma_{i_q,p}(\hat\bk_q)
\\
&&~~~~~~~~~~~~~~~~
\times\frac{a(|\bk_q|)}{|\bk_q|^{2\al+d-2}}\,\delta(\bk_q+\bk_{n+1})d\bk_q d\bk_{n+1}
\prod_{\widehat{rs}\in E(\cF)}
\Gamma_{i_r,i_s}(\hat\bk_r)\frac{a(|\bk_r|)}{|\bk_r|^{2\al+d-2}}\,\delta(\bk_r+\bk_s)d\bk_r d\bk_s.
\end{eqnarray*}
Using the change of variables made in the course of the proof of Proposition \ref{corr-est}
we conclude therefore that
\[
|\langle
\chi_{\la,p}^{(n)}({\cal F}),F_p\rangle_{L^2}|\le \frac{C}{\la^e}(1+|\log\la|)^{e+1},
\]
which yields \eqref{sc-pr} because $2e+1=n$.
$\Box$

\appendix
\section{Multiple stochastic integration}\label{sec:append}

Suppose that $\P$ is a Gaussian, homogeneous, Borel measure
over the space $\Om$, introduced in Section \ref{sec2.1}. Denote by
$\hat{\bF}$ the corresponding  Gaussian vector valued spectral measure
on $(\R^d,{\cal B}(\R^d))$. We suppose that the
structure measure $\hat{\bR}(d\bk)$ has the density ${\bf r}(\bk)=[r_{ij}(\bk)]$ w.r.t. the
Lebesgue measure. Let $r(\bk):=$tr ${\bf r}(\bk)\vee 1$.
For a given integer $n\ge1$ we consider the Borel
measure on $(\R^d)^{2n}$
$$
M_{2n}(d\bk_1,\ldots,d\bk_{2n}):=\sum\limits_{{\cal F}\in\mathfrak F_n}
\prod\limits_{\widehat{pq}\in {\cal F}}\delta(\bk_p+\bk_q)\prod_{j=1}^{2n}r(\bk_j)d\bk_j.
$$
By ${\cal L}_n^2$ we denote the completion of the space of all
complex valued, bounded
Borel measurable functions $\psi:(\mathbb R^d)^n\to(\mathbb C^d)^n$
in the norm
$$
\|\psi\|^2_{{\cal L}_n^2}:=\mathop{\int\ldots\int}\limits_{(\R^d)^{2n}}(|\psi(\bk_1,\ldots,\bk_n)|^2+
|\psi(\bk_{n+1},\ldots,\bk_{2n})|^2)M_{2n}(d\bk_1,\ldots,d\bk_{2n}).
$$
Suppose that  $i\in \Z^d$  and $N\ge1$.
 We define
\begin{equation}\label{a1}
\Box_{N}(i):=[\bk\in\R^d:\,2^{-N}i_{j}\le k_{j}<2^{-N}(i_{j}+1),\,\forall\,j=1,\ldots,d].
\end{equation}
For ${\bf i}:=( i_1,\ldots,i_n)\in (\Z^d)^n$ we let
\begin{equation}\label{a3}
\Box_{N}({\bf i}):=\Box_{N}(i_1)\times\ldots\times\Box_{N}(i_n).
\end{equation}
By ${\cal D}$ we denote the family of all such boxes.
Its  subfamily $\Pi$ is called \emph{an admissible dyadic  partition} of $(\R^d)^{2n}$
if

(P1) $\bigcup_{\Box\in \Pi}\Box=(\R^d)^{2n}$,

(P2) for any two boxes $\Box\not=\Box'\in\Pi$ we have $\Box\cap\Box'=\emptyset$,

(P3) there exists $d_0>0$ such that $|\Box|\ge d_0$ for all $\Box\in\Pi$.

A set function $c: \Pi\to (\mathbb C^d)^n$ is called \emph{admissible}
if $c(\Box)=0$ for all but finitely many boxes from an admissible $\Pi$.
We denote by ${\cal A}$ the family of all admissible set functions.

For any multi-index ${\bf j}:=( j_1,\ldots,j_n)\in \{1,\ldots,d\}^n$
and $\Box^{(N)}({\bf i})$ given by \eqref{a3}
we let
\begin{equation}\label{a2}
    \hat{\bF}_{\bf j}[\Box_{N}({\bf i})]=\prod\limits_{p=1}^n\hat{F}_{j_p}[\Box_{N}( i_p)].
\end{equation}
Suppose that $c: \Pi\to (\mathbb C^d)^n$ is admissible. We define then
$\psi(\bk):=c(\Box)$ for all $\bk\in\Box$. With some abuse of notation
we call such a function  admissible and
denote by ${\cal A}_n\subset {\cal L}_n^2$ the space of all admissible functions.
For any
admissible function $\psi\in{\cal A}_n$ we define the $n$--tuple stochastic
integral letting
\begin{equation}\label{a5}
{\cal I}(\psi):=
\sum\limits_{\Box\in \Pi}\sum\limits_{\bf j}c_{\bf j}(\Box)\cdot \hat{\bF}_{\bf j}[\Box].
\end{equation}
We shall also write $\int\!\ldots\!\int\psi(\bk_1,\ldots,\bk_n)\hat{\bF}(d\bk_1)\otimes\ldots\otimes\hat{\bF}(d\bk_n)$
to denote ${\cal I}(\psi)$.
Below, we  list some of the properties of
${\cal I}(\psi)$. They are elementary and their verification relies on
the application of the definition so we leave this task to a reader.
\begin{prop}
\label{propa1}

(i) ${\cal A}_n$ is dense in  ${\cal L}_n^2$ in the norm $\|\cdot\|_{{\cal L}_n^2}$.

(ii) The stochastic integral given by \eqref{a5} is well defined, i.e. if there exist
two admissible set functions $c_1,c_2$ corresponding to a given $\psi$ then
the respective definitions of the stochastic integrals are identical.

(iii) We have ${\cal I}(a_1\psi_1+a_2\psi_2)=a_1{\cal I}(\psi_1)+a_2{\cal I}(\psi_2)$

(iv) Suppose that $\psi_1,\ldots,\psi_n\in{\cal A}_1$. 
Then $\psi_1\otimes\ldots\otimes\psi_n\in{\cal A}_n$
and
$$
{\cal I}(\psi_1\otimes\ldots\otimes\psi_n)=\prod_{j=1}^n{\cal I}(\psi_j).
$$

(v) We have
\begin{eqnarray}
&&\E[{\cal I}(\psi^{(1)}){\cal I}^*(\psi^{(2)})]
=\sum\limits_{{\cal F}\in{\mathfrak F}_n}\sum\limits_{j_1,\ldots,j_{2n}=1}^d
\mathop{\int\ldots\int}\limits_{(\R^d)^{2n}}\psi^{(1)}_{j_1,\ldots,j_n}(\bk_1,\ldots,\bk_n)
(\psi^{(2)}_{j_{n+1},\ldots,j_{2n}})^*(\bk_{n+1},\ldots,\bk_{2n})\nonumber
\\
&&~~~~~~~~~~~~~~~~~~~~~~~
\times\prod\limits_{\widehat{pq}\in {\cal F}}\hat{R}_{pq}(\bk_p) \delta(\bk_p-\bk_q)
\prod_{j=1}^{2n}d\bk_j.\label{a10}
\end{eqnarray}

\end{prop}

As a direct consequence of property (v) and the definition of the norm on ${\cal L}_n$
we obtain.
\begin{cor}
\label{cora1}
 ${\cal A}_n\ni\psi\mapsto{\cal I}(\psi)$ is a continuous functional
 that extends to the entire ${\cal L}_n^2$. The extension shall be called
 an $n$-tuple stochastic integral w.r.t. the spectral measure $\hat{\bF}(d\bk)$.
 It has   properties (iii)-(v) from Proposition $\ref{propa1}$.
\end{cor}

\begin{prop}
\label{propa2}
For any $\psi\in{\cal L}_n^2$ we let
$$
U^\bx\psi(\bk_1,\ldots,\bk_n):=
\exp\left\{i\left(\sum_{p=1}^n\bk_p\right)\cdot\bx\right\}\psi(\bk_1,\ldots,\bk_n).
$$ 
Then, for any $\bx\in\R^d$ we have $U^\bx\psi\in{\cal L}_n^2$ and
\begin{equation}\label{a11}
    {\cal I}(\psi)(\tau_\bx\om)={\cal I}(U^\bx\psi)(\om),\quad\P-a.s.
\end{equation}
\end{prop}
\emph{Proof.}
We have
\[
\bF(\tau_{\bx+\by}\om)=\int e^{i\bk\cdot(\bx+\by)}\hat\bF(d\bk;\om)
=\bF(\tau_{\bx}(\tau_\by\om))=
\int e^{i\bk\cdot\bx}\hat\bF(d\bk;
\tau_\by\om)
\]
for all $\bx,\by\in\R^d$. Hence $\hat\bF(d\bk;
\tau_\by\om)=e^{i\bk\cdot\by}\hat\bF(d\bk;
\om)$ and \eqref{a11} follows.
$\Box$

{\bf Remark} A different construction of a multiple stochastic integral
can be obtained using the approach of Chapter 7.2 of \cite{janson}.
Denote the stochastic integral, defined there, by
$$\int\!\ldots\!\int^{(J)}\psi(\bk_1,\ldots,\bk_n)\cdot
\hat{\bF}(d\bk_1)\otimes\ldots\otimes\hat{\bF}(d\bk_n).$$
Let  ${\cal F}$  be a Feynman diagram labelled
by $\{1,\ldots,n\}$. Let $n_1<\ldots< n_a$ be all the free vertices of $\cF$
and let $\widehat{pq}\in E(\cF)$
denote the remaining $e:=(n-a)/2$ edges.
Set $T(\cF)\psi:(\mathbb R^d)^a\to(\mathbb C^d)^a$
\[
T(\cF)\psi_{i_{n_1}\ldots i_{n_a}}(\bk_{n_1},\ldots,\bk_{n_a})
:=\sum_{i_pi_q}\int\ldots\int
\psi_{i_{1}\ldots i_{n}}(\bk_{1},\ldots,\bk_{n})\prod_{\widehat{pq}\in E(\cF)}r_{i_p i_q}(\bk_p)
\delta(\bk_p+\bk_q)d\bk_p d\bk_q
\]
for all $i_{n_1},\ldots, i_{n_a}=1,\ldots,d$.
Let
$$
I({\cal F}):=\int\!\ldots\!\int^{(J)}T(\cF)\psi(\bk_{n_1},\ldots,\bk_{n_l})
\hat{\bF}(d\bk_{n_1})\otimes\ldots\otimes\hat{\bF}(d\bk_{n_l}).
$$
Using Theorem 7. 25 p. 99
and Corollary 3. 17 p. 28 of ibid. one can conclude that
\begin{equation}\label{a21}
\int\!\ldots\!\int\psi(\bk_1,\ldots,\bk_n)\hat{\bF}(d\bk_1)\otimes\ldots\otimes\hat{\bF}(d\bk_n)
=\sum_{\cal F}I({\cal F}),
\end{equation}
where the summation extends over all Feynman diagrams labelled
by $\{1,\ldots,n\}$.
We could use then equality \eqref{a21} as the definition of the multiple stochastic integral.

\end{document}